\documentclass[aps,amsmath,amssymb,twocolumn,longbibliography,showkeys,10pt]{revtex4-1}
\usepackage[T1]{fontenc}
\usepackage{graphicx}
\usepackage[colorlinks=true,linkcolor=blue,citecolor=blue,urlcolor=blue]{hyperref}
\usepackage{charter}

\usepackage{framed}
\usepackage{wrapfig}

\begin{document}   

\title{Simulating artificial one-dimensional physics \\ with ultra-cold fermionic atoms: three exemplary themes}
\author{Jacek Dobrzyniecki}
\affiliation{ 
Institute of Physics, Polish Academy of Sciences, Aleja Lotnik\'ow 32/46, PL-02668 Warsaw, Poland}
\author{Tomasz Sowi\'nski}
\email{tomsow@ifpan.edu.pl}
\affiliation{
Institute of Physics, Polish Academy of Sciences, Aleja Lotnik\'ow 32/46, PL-02668 Warsaw, Poland}
\date{\today}

\begin{abstract}
For over twenty years, ultra-cold atomic systems have formed an almost perfect arena for simulating different quantum many-body phenomena and exposing their non-obvious and very often counterintuitive features. Thanks to extremely precise controllability of different parameters they are able to capture different quantum properties which were previously recognized only as theoretical curiosities. Here, we go over the current experimental progress in exploring the curious one-dimensional quantum world of fermions from the perspective of three subjectively selected trends being currently under vigorous experimental validation: ({\it i}) unconventional pairing in attractively interacting fermionic mixtures, ({\it ii}) fermionic systems subjected to the artificial spin-orbit coupling, ({\it iii}) fermionic gases of atoms with high $\mathrm{SU}({\cal N})$ symmetry of internal states.
\end{abstract} 
\maketitle
\begin{framed}
\begingroup
\setlength{\columnsep}{14pt}%
\setlength{\intextsep}{5pt}%
\begin{wrapfigure}{l}{0.4\textwidth}
\includegraphics[width=1\linewidth]{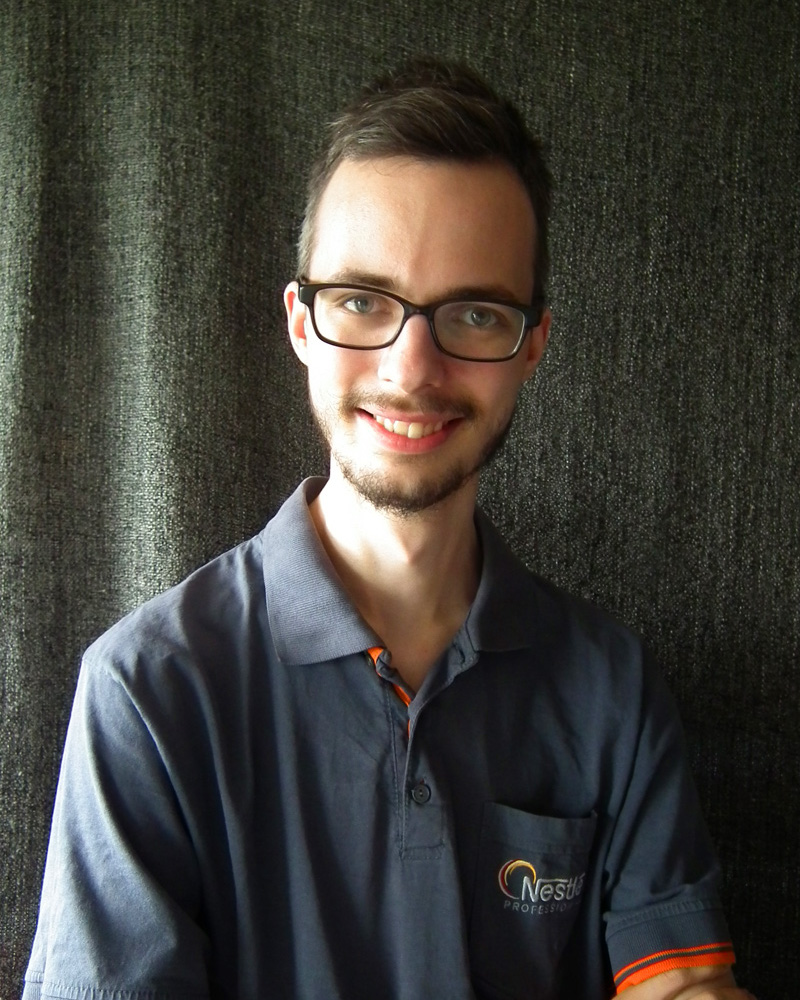}
\end{wrapfigure}
\textbf{Jacek Dobrzyniecki} is a last-year Ph.D. student at the Institute of Physics, Polish Academy of Sciences, Warsaw. The main focus of his research is static and dynamic properties of few-atom ultra-cold systems in one dimension. His scientific interests include ultra-cold atomic physics, strongly correlated few-body systems, low-dimensional quantum systems, and open quantum systems.

\endgroup
\begingroup
\setlength{\columnsep}{14pt}%
\setlength{\intextsep}{5pt}%
\begin{wrapfigure}{l}{0.4\textwidth}
\includegraphics[width=1\linewidth]{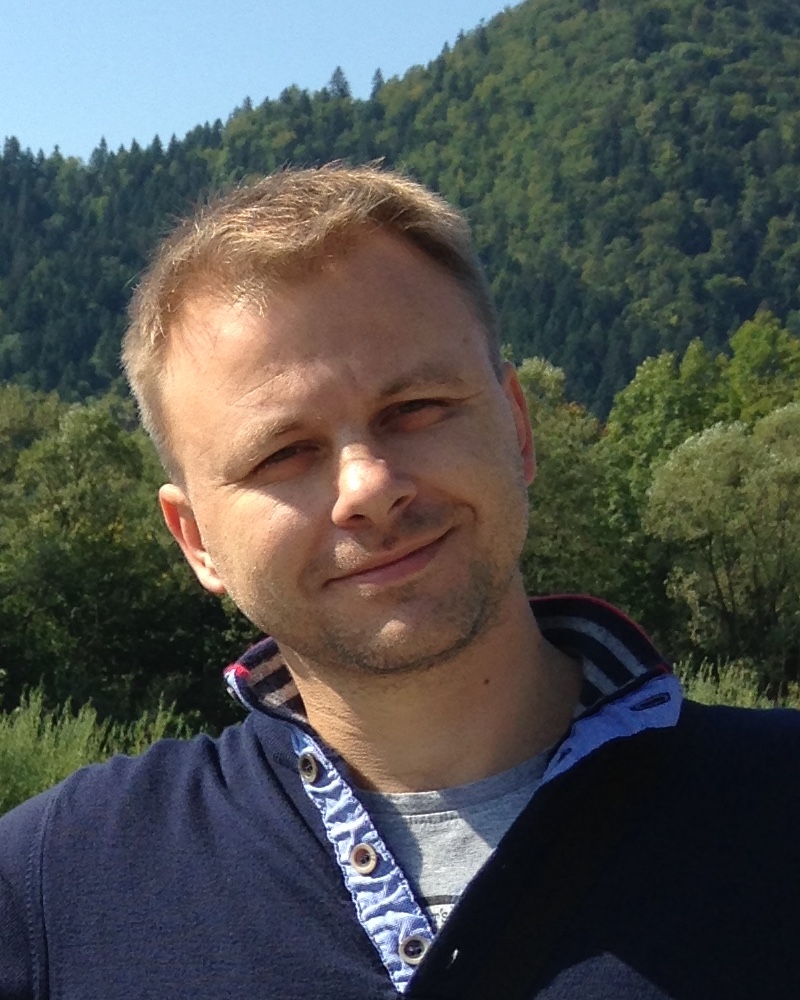}
\end{wrapfigure}
\textbf{Tomasz Sowi\'{n}ski} received his Ph.D. in physics in 2008 from the University of Warsaw. Currently, he is an associate professor at the Division of Theoretical Physics of the Institute of Physics, Polish Academy of Sciences, Warsaw, where he leads The Few-Body Problems Theory Group. His scientific interests focus mostly on ultra-cold atomic physics, quantum simulations, quantum systems with reduced dimensionality, and strongly correlated few-body systems.

\endgroup
\end{framed}
\tableofcontents

\section{Introduction}
Physical sciences were born mainly to deliver an appropriate and understandable description of the observable world. Typically, the laws of physics were formulated {\it after} discovery of related phenomena and eventually then their general consequences were disclosed. However, from time to time, some artificial theoretical models (frequently additionally reduced to one spatial dimension) were introduced without any experimental motivation -- just to expose some intriguing properties of the underlying theory \cite{1968CooperBook}. This kind of approach was intensified when quantum mechanics was born since then many realistic problems were considered as too complicated to be explained in a full quantum-mechanical treatment. This is how many interesting simplified theoretical models were introduced. Let us mention here only a few celebrated examples: the Dicke model \cite{1954DickePR,1973HeppAnnPhys}, the Schwinger model \cite{1962SchwingerPR}, the Hubbard model \cite{1963HubbardProcRSocLondon}, the Jaynes-Cummings model \cite{1963JaynesProcIEEE}, the Lieb-Liniger model \cite{1963LiebPhysRev1,1963LiebPhysRev2}, the Calogero-Sutherland model \cite{1971CalogeroJMP,1971SutherlandJMP}, the Harper-Hofstadter lattice model \cite{1976HofstadterPRB}, the Sachdev-Ye-Kitaev model \cite{1993SachdevPRL}, the Kitaev chain model \cite{2001KitaevPhysUsp}, the toric code model \cite{2003KitaevAnnPhys}, the Kitaev honeycomb model \cite{2006KitaevAnnPhys} {\it etc.} The situation changed drastically along with tremendous development of experimental methods of precisely controlling interactions between light and matter on a subatomic scale. It turns out that these simplified theoretical models and a variety of their extensions and modifications are possible to engineer with atomic systems, {\it i.e.}, appropriately prepared and controlled ultra-cold atomic systems may serve as almost perfect realizations of systems described by desired Hamiltonians \cite{LewensteinBook}. In this way, Richard Feynman's brilliant dream of creating {\it quantum simulators} \cite{1982FeynmanIJTP} can be realized and a new era of {\it quantum engineering} has started. Importantly, such quantum simulators not only can verify many theoretical predictions, but also shed light on long-standing questions that have not been satisfactorily answered by theoretical analysis.

One example of the exciting developments in the field is the ability to engineer effectively one-dimensional quantum many-body systems. This idea has long been of theoretical interest, including such important models as the Tonks model of a gas of impenetrable rods \cite{1936TonksPR} or the Gaudin-Yang general solution for the ground state of fermions \cite{1967GaudinPLA,1967YangPRL}. Now, by utilizing the quantum simulator concept, many of the exotic physical phenomena characterizing one-dimensional systems can be explored experimentally for the first time. With advanced trapping techniques, atoms can be confined in traps of effectively varying dimensionality by controlling the strength of perpendicular confinement. This includes optical lattices \cite{2005BlochNatPhys} and single- and few-site potentials \cite{2011CazalillaRevModPhys,2013GuanRevModPhys,2019SowinskiRepProgPhys}. In particular, different one-dimensional systems of ultra-cold fermionic mixtures have been experimentally created in this way \cite{2005MoritzPRL,2010LiaoNature}. 

In this review, we describe recent achievements in the domain of one-dimensional fermionic ultra-cold atom systems. Our focus is on the developments that have occurred in the past few years, since the last comprehensive review from 2013 by Guan \emph{et al} \cite{2013GuanRevModPhys}. Since the progress in the entire field of 1D fermionic systems is exceedingly broad and rapid, a full catalogue of all the major advancements would be a tremendous undertaking. Therefore we concentrate on three subjectively selected main research directions which are currently being heavily explored and, in our opinion, will have significant importance for the future capability of quantum technologies. 

The organization of this article is as follows. In Section \ref{sec-fflo} we describe the search for unconventional superconducting states in one-dimensional systems, which are of considerable theoretical interest but are still difficult to pin down experimentally. In Section \ref{sec-soc} we describe systems under the influence of artificial spin-orbit coupling. In one-dimensional settings, this kind of coupling presents an interesting picture, since there is in fact no ``orbit'' in the usual sense. Furthermore, it has important applications, such as the simulation of topologically nontrivial models requiring higher dimensionality. In Section \ref{sec-sun} we describe the research on atomic systems with higher-spin internal symmetries, which in the case of one spatial dimension offer a fascinating arena for exploring various exotic many-body phases. Section \ref{sec-concl} is the conclusion.


\section{Unconventional pairing phases}
\label{sec-fflo}

The simplest properties of superconducting materials are typically described by the pairing mechanism of spontaneous formation of correlated pairs of opposite-spin fermions. The mechanism is appropriately described by the theory of superconductivity of Bardeen, Cooper, and Schrieffer (BCS) \cite{1957BardeenPR}. However, when there is no direct symmetry between opposite-spin components, certain more exotic pairing phases are possible. One of them, the \emph{Fulde-Ferrell-Larkin-Ovchinnikov} (FFLO) pairing phase, has recently attracted significant interest.

The FFLO phase was first predicted in the 1960s by Fulde and Ferrell \cite{1964FuldePR} and independently by Larkin and Ovchinnikov \cite{1965LarkinSoviet}, who considered the ground state of a solid-state superconductor subjected to an external magnetic field. The magnetic field causes a relative shift of the Fermi surfaces of both electron spin components. If this shift is too high, the Cooper pairing is destroyed and the transition from the superconducting to the normal state occurs. However, Fulde, Ferrell, Larkin, and Ovchinnikov showed that close to the transition (still in superconducting phase) an FFLO state can be formed. In this unusual region the pairing of fermions with momenta $\vec{k}$ and $-\vec{k}+\vec{Q}$ is favored over the standard BCS pairing of fermions with momenta $\vec{k}$ and $-\vec{k}$ \cite{2004CasalbuoniRevModPhys}. The resulting Cooper pairs have a nonzero center-of-mass momentum $\vec{Q}$ which is (in general) proportional to the magnitude of the mismatch between the Fermi wave vectors \cite{2018KinnunenRepProgPhys}. A signature feature of the FFLO phase is that the pairing order parameter $\Delta(\vec{x})$ is no longer constant in space as in the standard BCS phase, but rather has an oscillatory character, $\Delta(\vec{x}) \propto \cos(\vec{Q} \cdot \vec{x})$ \cite{2007MatsudaJpn}. 

In the decades since the original proposals, the FFLO phase has been extensively investigated. It has been the subject of several reviews \cite{2004CasalbuoniRevModPhys,2007MatsudaJpn,2013BeyerLowTemp}, including very recent ones \cite{2018KinnunenRepProgPhys,2018WosnitzaAnnPhys,2020AgterbergAnnRevCondMat}. The FFLO state is currently invoked to explain the behavior of several superconducting systems, including heavy-fermion and organic materials, as well as the cores of neutron stars \cite{2001AlfordJPhysG,2004CasalbuoniRevModPhys}. However, in spite of its significance, universally accepted experimental evidence for the FFLO pairing has still not been obtained, although a number of experiments conducted in solid-state systems have shown results highly suggestive of the FFLO state \cite{2007MatsudaJpn,2017CroitoruCondMat,2018WosnitzaAnnPhys,2003RadovanNature,2008YonezawaPRL,2008YonezawaJpn,2017ChoPRL,2018KitagawaPRL,2020KasaharaPRL}. Quasi-one-dimensional ultra-cold fermionic systems with attractive interactions offer another promising route to its experimental demonstration.

\subsection{Quantum simulators in one-dimensions}

Lower-dimensional systems, such as (quasi-)one-dimensional systems, are highly preferred in the experimental search for the FFLO phase. For three-dimensional fermionic systems, mean-field theoretical results indicate that the FFLO state is very unstable, and may exist only in a tiny sliver in the phase diagram \cite{2006SheehyPRL,2007SheehyAnnPhys,2007ParishNatPhys}. On the other hand, in one-dimensional systems the FFLO phase occupies a significant portion of the phase diagram \cite{1987BuzdinSoviet,2007HuPRL,2007ParishPRL,2012FeiguinChapter}. One major reason for this difference is that the stability of the FFLO state depends on the nesting between the Fermi surfaces. The simplest argument comes from the observation that, in lower dimensions, the number of states with total momentum $|\vec{Q}|$ is significantly reduced and therefore condensation of pairs to the FFLO state with a particular $\vec{Q}$ is facilitated. Of course, the rigorous picture is more complicated, but still the effect of dimensionality is crucial \cite{2017PtokJPhys,2018KinnunenRepProgPhys,2018ItahashiJPhysJapan}. Another important effect arises for systems of charged particles (such as solid-state superconductors). Applying an external magnetic field to the charged particles typically causes orbital effects, which is destructive to superfluidity. However, in lower-dimensional systems this detrimental effect is suppressed due to geometric constraints \cite{2007MatsudaJpn,2014MiyawakiJPhys,2018KinnunenRepProgPhys}. As a result, (quasi-)one-dimensional systems are a good environment to search for the elusive FFLO state. 

For this reason, the unconventional FFLO pairing in ultra-cold one-dimensional systems has recently been deeply investigated theoretically from various perspectives, for both confined \cite{2007OrsoPRL,2008CasulaPRA,2019PecakPRR,2019LydzbaPRA} and lattice systems \cite{2007FeiguinPRB,2010HeidrichPRA,2012FrancaPRA,2017FrancaPhysA,2018DePicoliBrazJPhys}. Quasi-one-dimensional quantum simulators created with ultra-cold neutral atoms constitute a highly controllable environment, where the Fermi surface mismatch can be precisely tuned by changing the spin composition of the initial population, rather than with external magnetic fields \cite{2006ZwierleinScience,2006PartridgeScience,2006ZwierleinNature,2006ShinPRL,2007SchunckScience}. The relative spin populations can be tuned, for example, by driving radio-frequency sweeps between the states at different powers \cite{2010LiaoNature}. 

\begin{figure}
\includegraphics[width=1\linewidth]{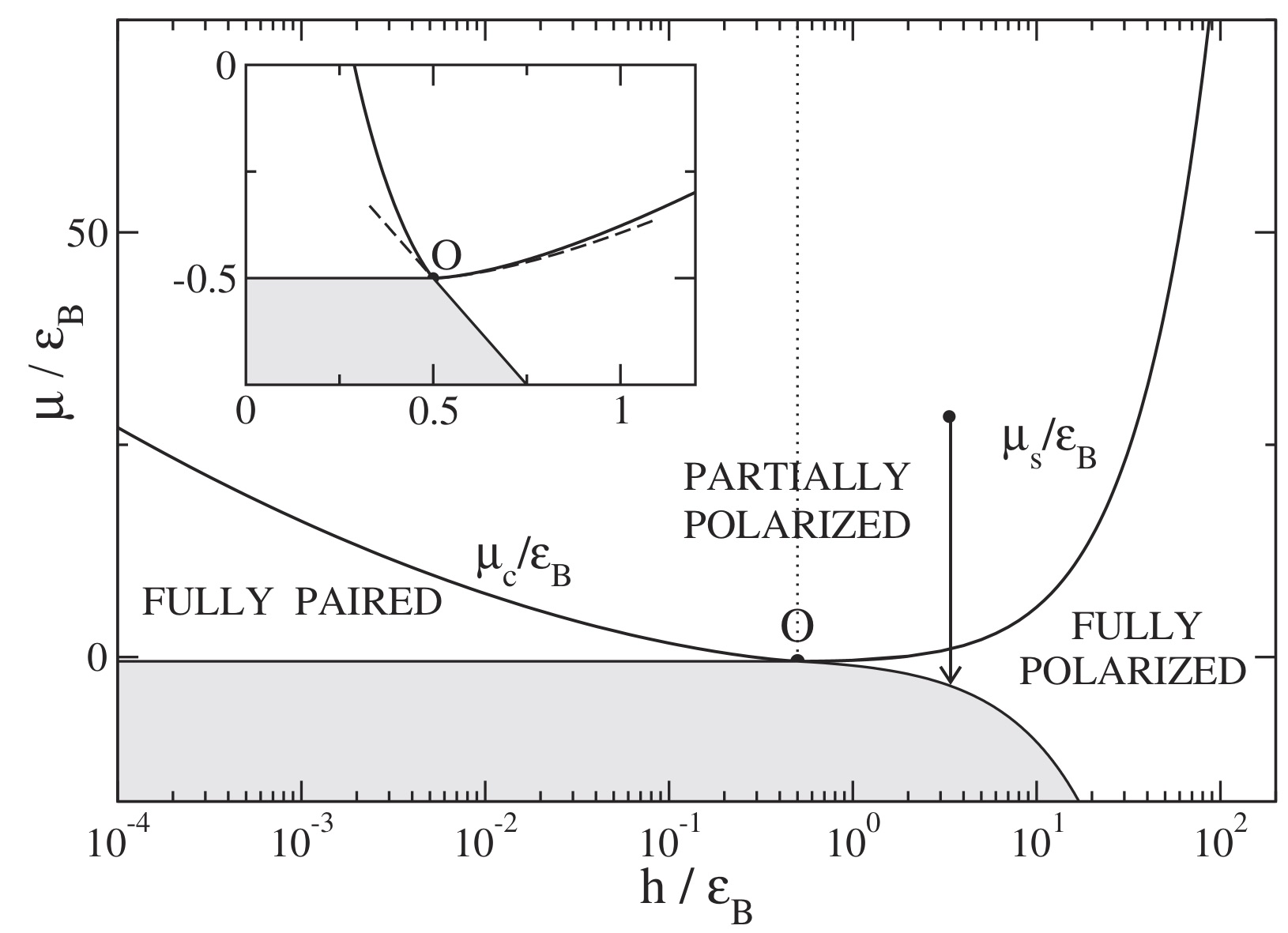}
\caption{The ground state phase diagram of an attractive homogeneous Fermi gas, in the plane of average chemical potential $\mu = (\mu_\uparrow + \mu_\downarrow)/2$ vs. effective magnetic field $h = (\mu_\uparrow - \mu_\downarrow)/2$, where $\mu_\uparrow,\mu_\downarrow$ are the chemical potentials of individual spin components. $\mu$ and $h$ are given in units of binding energy $\epsilon_B$. $\mu_c$ and $\mu_s$ designate the critical values of $\mu$ that separate the distinct phases. One can distinguish between the ``partially polarized'' phase (i.e. the FFLO-type phase), the ``fully paired'' phase (standard BCS Cooper pairing), and the ``fully polarized'' phase (unpaired state). The shaded region corresponds to the vacuum state. The arrow shows the trajectory of local chemical potential for a system in a harmonic trap, for which the local value of $\mu(x)$ decreases away from the trap center. Inset: zoom of the phase diagram near the point $O = (\epsilon_B/2,-\epsilon_B/2)$, with asymptotic behavior of phase diagram boundaries marked with dashed lines. Reproduced with permission from  \cite{2007OrsoPRL}. Copyright 2007, American Physical Society. 
 \label{Fig1} }
\end{figure}

One of the simplest models for such a one-dimensional system is that of a homogeneous Fermi gas with attractive contact interactions \cite{2007OrsoPRL}. It can be described by the Gaudin-Yang Hamiltonian of the form
\begin{equation}
 H = - \frac{\hbar^2}{2m}\sum\limits_{i=1}^{N_\downarrow+N_\uparrow} \frac{\partial^2}{\partial x_i^2} + g_{1D}\sum\limits_{i=1}^{N_\uparrow}\sum\limits_{j=1}^{N_\downarrow}\delta(x_i-x_j),
\end{equation}
where $N_\sigma$ is the number of fermions with spin $\sigma \in \{\uparrow,\downarrow\}$, $m$ is the fermion mass, $x_i$ is the position of the $i$-th fermion and $g_{1D}$ is the strength of the contact interaction (attractive for $g_{1D} < 0$). In the large particle number limit, one can define the chemical potentials of individual spin components, $\mu_\sigma = \partial E / \partial n_\sigma$, as the derivatives of system energy $E$ over the density $n_\sigma$ of the given component. Shown in Fig.~\ref{Fig1} is the theoretical phase diagram of this system at zero temperature, in the plane of the chemical potential, $\mu = (\mu_\uparrow + \mu_\downarrow)/2$, and the spin population imbalance, $h = (\mu_\uparrow - \mu_\downarrow)/2$ (which is equivalent to the strength of the effective magnetic field). At low spin imbalance, the ground state of this system is the standard BCS paired phase. When the imbalance is increased, the system transitions into the FFLO-paired phase. At a high enough imbalance, the superfluid phase is destroyed and the system is in the normal, unpaired phase. 

\begin{figure}
\includegraphics[width=1\linewidth]{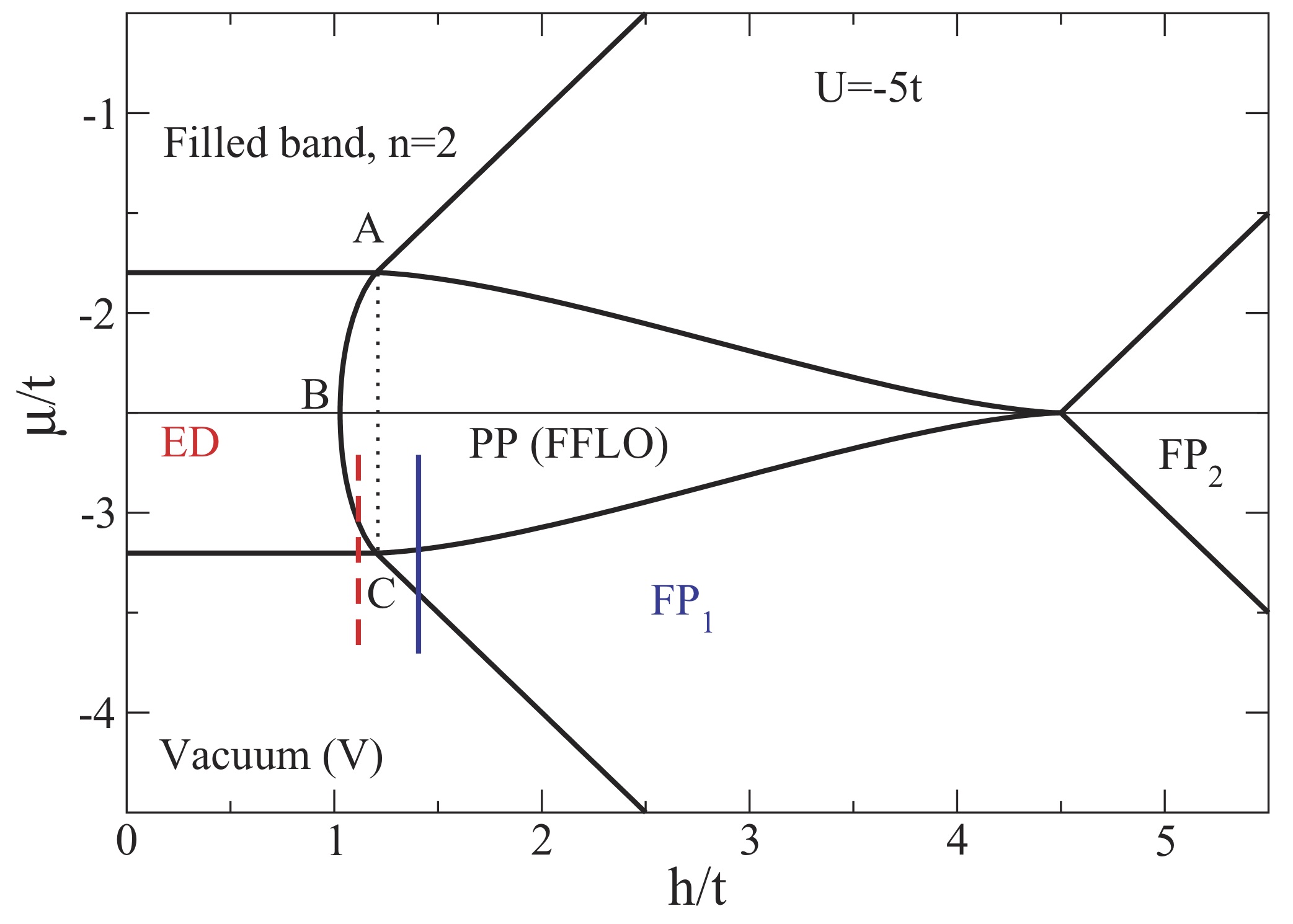}
\caption{Grand canonical phase diagram of the one-dimensional attractive Hubbard model (for fixed interaction $U=-5t$) in the plane of average chemical potential $\mu$ vs. effective magnetic field $h$. $\mu$ and $h$ are given in units of tunneling amplitude $t$. The distinct phases visible are as follows: V -- vacuum, ED -- unpolarized phase (standard BCS pairing), PP -- partially polarized phase (FFLO-type pairing), FP$_1$ -- fully polarized (unpaired) phase with low filling $n < 1$, FP$_2$ -- fully polarized (unpaired) phase with filling $n = 1$. The remaining phases correspond to the states where at least one of the components forms a band insulator. When this system is subject to additional harmonic confinement, the local chemical potential $\mu(x)$ becomes smaller as one goes from the trap center to the edges, for example as shown by the two trajectories (red and blue lines). Reproduced with permission from \cite{2010HeidrichPRA}. Copyright 2010, American Physical Society. 
}
 \label{Fig2} 
\end{figure}

Fig.~\ref{Fig2} shows an analogous phase diagram for a 1D lattice system in the tight-binding approximation \cite{2010HeidrichPRA}, described by the Hubbard-like model
\begin{equation}
 H = \sum_j\left[\sum_\sigma -t\,\hat{c}^\dagger_{j,\sigma} ( \hat{c}_{j-1,\sigma} + \hat{c}_{j+1,\sigma}) + U  \hat{n}_{j,\uparrow} \hat{n}_{j,\downarrow}\right],
\end{equation}
where $t$ is the hopping amplitude between neighboring sites, $U$ is the on-site interaction energy, $\hat{c}_{i,\sigma}$ is the annihilation operator for a fermion with spin $\sigma$ at site $i$, and $\hat{n}_{i,\sigma} = \hat{c}^\dagger_{i,\sigma} \hat{c}_{i,\sigma}$. Despite the differences between the two systems, the overall structure of this phase diagram is similar to the homogenous 1D gas case, with the standard BCS-paired phase transitioning to the FFLO phase at a finite spin imbalance, and a subsequent transition to the normal phase beyond a critical imbalance value. 

One should remember that in the presence of external trapping, due to inhomogeneity, different configurations predicted by these phase diagrams may simultaneously co-exist at different locations in the trap. This phase separation can be understood in the framework of the local density approximation picture, which is useful when the particle density varies slowly in space. In this approach, the local value of the chemical potential $\mu(x)$ varies along the system as $\mu(x) = \mu^0 - V(x)$, where $\mu^0$ is the chemical potential at the center of the trap and $V(x)$ is the trap potential \cite{2010HeidrichPRA}. In particular, for a harmonic trap, $\mu(x)$ becomes smaller towards the trap edges. Meanwhile, the effective magnetic field $h(x)$ remains constant throughout the trap. The result is that different phases are realized at different locations, and their arrangement corresponds to a trajectory across the phase diagram, starting at $\mu^0$ and going downwards parallel to the $\mu$ axis. Examples can be seen in Fig.~\ref{Fig1} and Fig.~\ref{Fig2}, where the trajectories of the local chemical potential across the trap length are shown as vertical lines which pass through several different phases, indicating the presence of different phases at different locations in the trap. For a harmonically trapped 1D spin-imbalanced gas (whether with or without a lattice), a typical configuration is a two-shell structure, where the center of the system exhibits an FFLO phase and the edges of the system are in the unpaired phase or the standard BCS phase \cite{2007FeiguinPRB,2007OrsoPRL,2010HeidrichPRA}. 

Depending on the structure of the phase diagram, more complex phase-separation configurations can be obtained. This possibility was explored in \cite{2019CichySciRep}, where it was shown how, by appropriately modifying the parameters of the confining trap, one can engineer different trajectories in the phase diagram. In this way, a desired configuration of separated phases can be created. 

\begin{figure}
\includegraphics[width=1\linewidth]{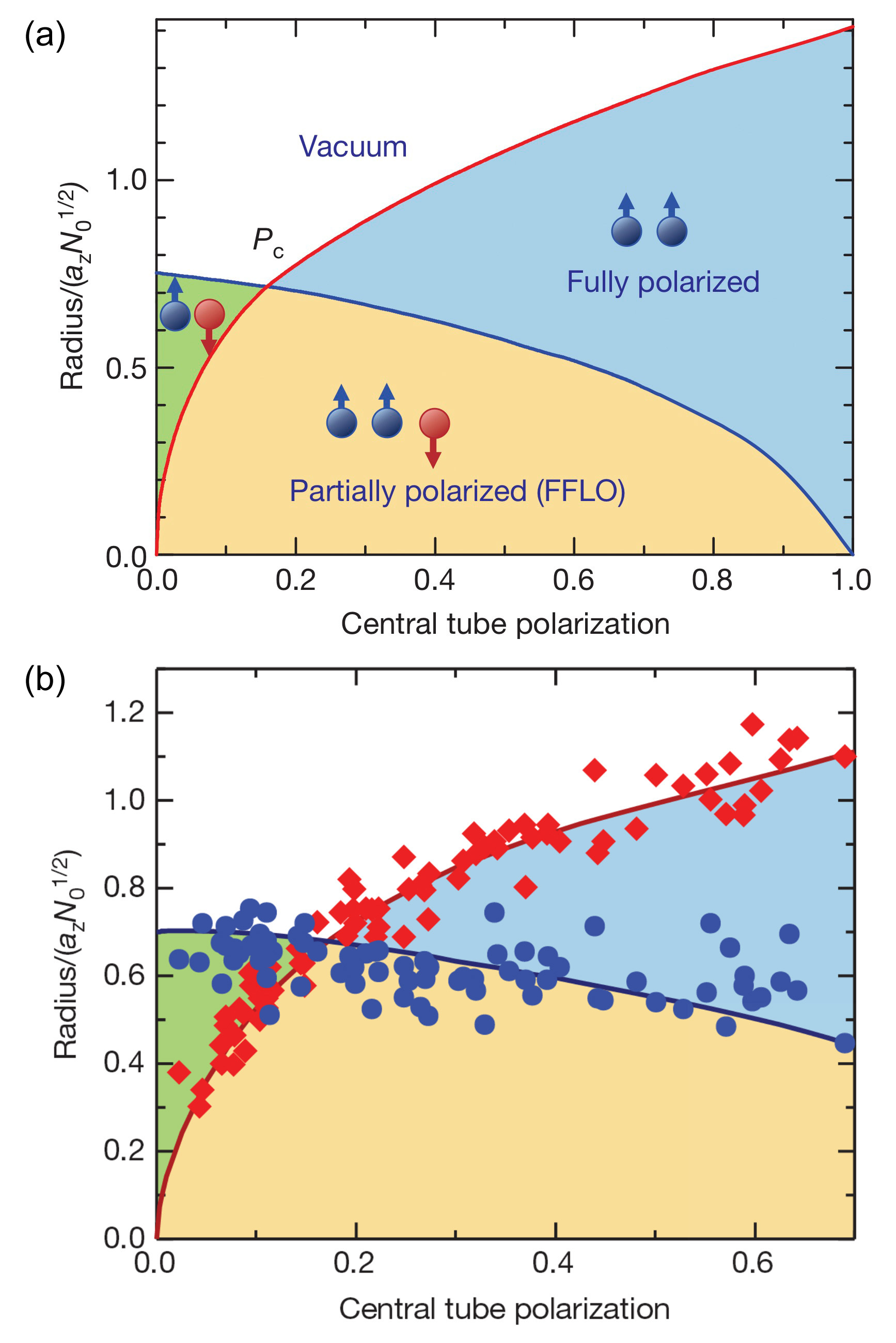}
\caption{(a) Theoretical phase diagram of a homogeneous Fermi gas trapped in an array of 1D tubes at zero temperature, as a function of the spin polarization of the central 1D tube and the radius from the center of the tube. The red and blue lines indicate the radii at which boundaries between the different phases are predicted to occur. $P_c$ is the critical polarization value where the boundaries cross. The distinct phases shown are: the vacuum (white), fully paired (standard BCS-type) phase (green), partially polarized (FFLO-type) phase (orange), and the fully polarized (unpaired) phase (blue). (b) Comparison of theoretical predictions (red and blue solid lines) and the experimentally measured locations of the boundaries (red and blue symbols) at temperature 175 nK. Reproduced with permission from \cite{2010LiaoNature}. Copyright 2010, Nature.}
 \label{Fig3} 
\end{figure}

This theoretical phase separation picture was confirmed experimentally by the group of Randall G. Hulet and Erich J. Mueller \cite{2010LiaoNature}. The experiment studied the nature of phase separation which occurred in fermionic systems confined to 1D tubes as the total spin imbalance of the system was tuned. Fig.~\ref{Fig3}a depicts the theoretically predicted locations of the boundaries (red and blue lines) between different phases, shown in terms of the radius from the trap center. The locations of the boundaries between the phases, and the particular phases realized, can be seen to depend on the value of the spin polarization. Fig.~\ref{Fig3}b compares these theoretical predictions with the experimentally measured locations of the boundaries (red and blue symbols). At low spin polarization, below a critical value $P_c$, the center of the cloud was occupied by a partially polarized state. Towards the edges of the cloud, the system was fully paired. For values of polarization close to $P_c$, the partially polarized phase extended across the entire trap. Finally, at high polarization $P > P_c$ the state at the edge of the cloud changed from fully paired to fully polarized, in agreement with the theoretical prediction. 

Undoubtedly, the experiment showed the validity of the predicted phase separation. Nevertheless, it did not provide direct evidence that the partially polarized state in the center of the trap was indeed the elusive FFLO state. In anticipation of future experimental work, several potential experimental signatures of the FFLO state have been proposed. A well-established possibility is measuring the pair momentum distribution of the partially polarized phase, with a peak at finite momentum $q$ providing an unambiguous signature of FFLO with pair momentum $q$ \cite{2008CasulaPRA,2010LiaoNature}. In recent years, there have been proposals based on surveying the expansion dynamics of the cloud after switching off the trapping potential \cite{2011KajalaPRA,2012LuPRL}. Recently, it was also suggested \cite{2016YinPRA} that the FFLO state subject to a sudden quench of the interaction strength should display characteristic, experimentally detectable post-quench features. It has also been shown that the visibility of the FFLO state should be greatly enhanced in one-dimensional boson-fermion mixtures with strong boson-fermion repulsion \cite{2019Singh}. 

\subsection{Dimensionality crossover}


For future experimental work, promising perspectives are opened by the implementation of systems with an ``intermediate'' dimensionality. Although here we focus on 1D systems, the dimensional crossover technique offers interesting perspectives for the observation of FFLO and it is worth looking at recent developments in this area. Experimentally, a system of this kind can be implemented with an array of 1D tubes, where the amplitude of tunneling between neighboring tubes $t_\perp$ can be tuned, thus controlling the effective dimensionality. In particular, in lattice systems the relevant parameter is the ratio $t_\perp/t_\parallel$ (where $t_\parallel$ is the tunneling between sites of a single tube), which can range from 0 (fully 1D) to 1 (isotropic 3D) \cite{2012KimPRB,2013HeikkinenPRB}. A quasi-1D regime, where $t_\perp$ is small but nonzero, is expected to be even better suited to the observation of the FFLO state than a purely 1D system. It comes from the fact that, in such a regime, there can exist long-range order absent from purely 1D systems, stabilizing the FFLO phase \cite{2007ParishPRL,2013HeikkinenPRB}. Furthermore, correlations induced by the weak intertube tunnelings could synchronize the FFLO density modulations across different tubes and thus enhance the overall experimental signal from the array \cite{2007ParishPRL}. 

The difference between different dimensionalities manifests itself in the phase separation of a trapped Fermi gas. As noted above, in a quasi-one-dimensional trap one typically obtains a two-shell structure where the center of the trap is occupied by a partially polarized FFLO state while the edges are taken up by a polarized normal state. On the other hand, in a spherical 3D trap, a shell structure is predicted with the standard BCS superfluid occupying the trap center \cite{2007LiuPRA-Mean}. Theoretical and experimental research has confirmed that the crossover between 1D and 3D dimensionalities is indeed reflected in the phase separation structure of the system. 

\begin{figure}
\includegraphics[width=1\linewidth]{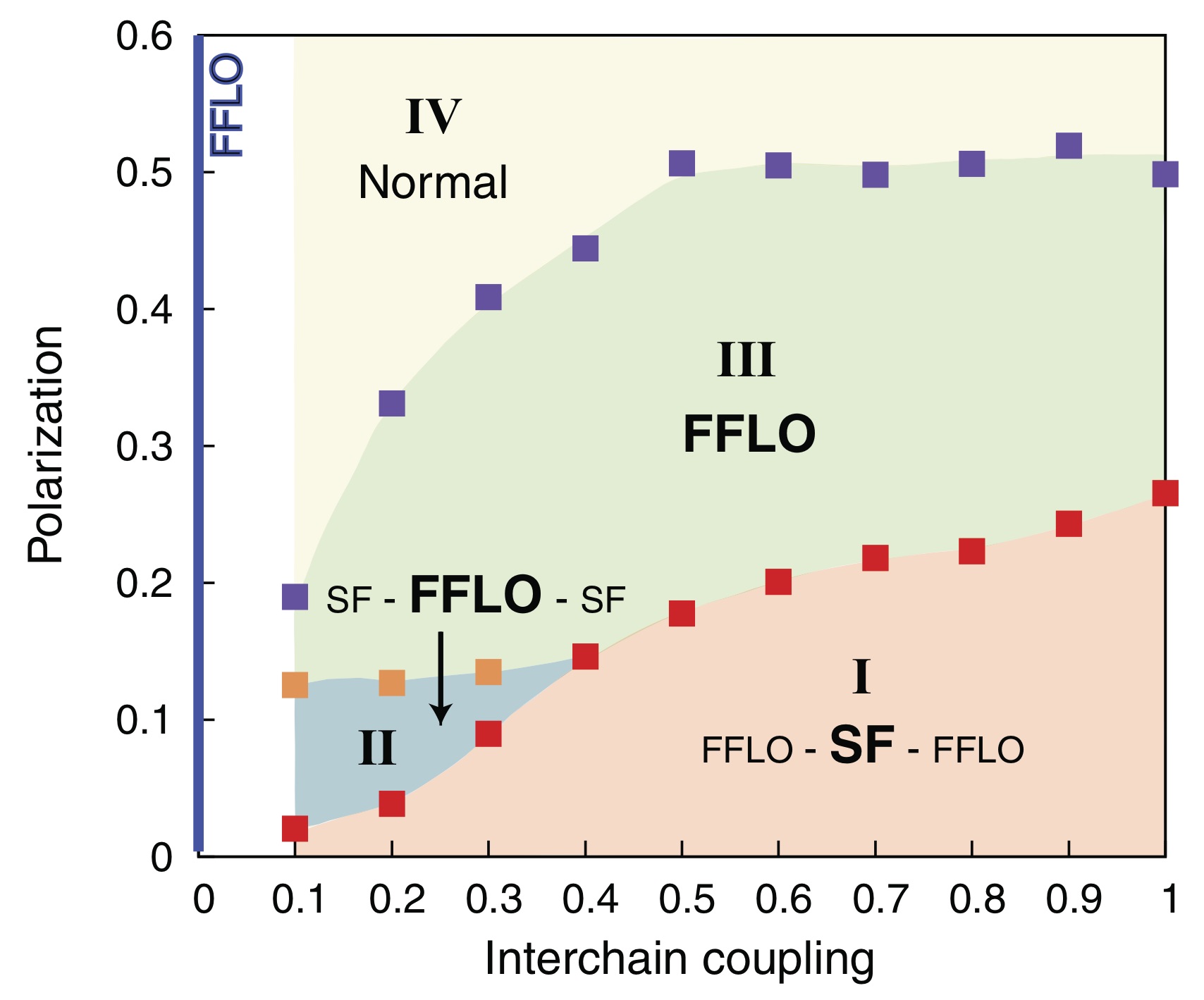}
\caption{Phase diagram of the model representing an array of coupled one-dimensional Hubbard chains, in the plane of interchain coupling vs. spin polarization. Interchain coupling is given as the ratio $t_\perp/t_\parallel$ between transverse and inter-site tunnelings. Phase III represents a two-shell structure, with an FFLO core and fully polarized, unpaired edges. Phase II represents a three-shell structure with an FFLO core, standard BCS superfluid in the shoulders, and fully polarized edges. Phase I represents a three-shell structure with a standard BCS superfluid core, FFLO shoulders, and fully polarized edges. Reproduced with permission from \cite{2012KimPRB}. Copyright 2012, American Physical Society.}
 \label{Fig4} 
\end{figure}

For example, in \cite{2012KimPRB,2013HeikkinenPRB} the phase diagram of a 3D array of 1D lattices with harmonic trapping was studied. The zero-temperature phase diagram of the relevant system can be seen in Fig.~\ref{Fig4}. At $t_\perp/t_\parallel > 0$, in addition to the familiar two-shell structure with FFLO in the center (region ``III'' in Fig.~\ref{Fig4}), there appears the possibility of obtaining three-shell structures, in which the gas in the center of the trap separates into two shells displaying FFLO and standard BCS phases (regions ``I'' and ``II'' in Fig.~\ref{Fig4}). As the transverse coupling increases, the structure with standard BCS pairing occupying the trap center (region ``I'' in the trap center), characteristic of a quasi-three-dimensional regime, becomes preferred \cite{2012KimPRB}. It is argued that the approximate crossover point between quasi-one-dimensional and quasi-three-dimensional physics, $t_\perp/t_\parallel \approx 0.3$, is a ``sweet spot'' where the FFLO state displays a highly uniform oscillation amplitude across the entire 1D tube \cite{2013HeikkinenPRB}. Above a critical temperature, which is approximately 1/3 that of the critical temperature for BCS superconductivity, the FFLO phase becomes fragile to losing its FFLO character and melting into standard BCS pairing \cite{2013HeikkinenPRB}. 

\begin{figure}
\includegraphics[width=1\linewidth]{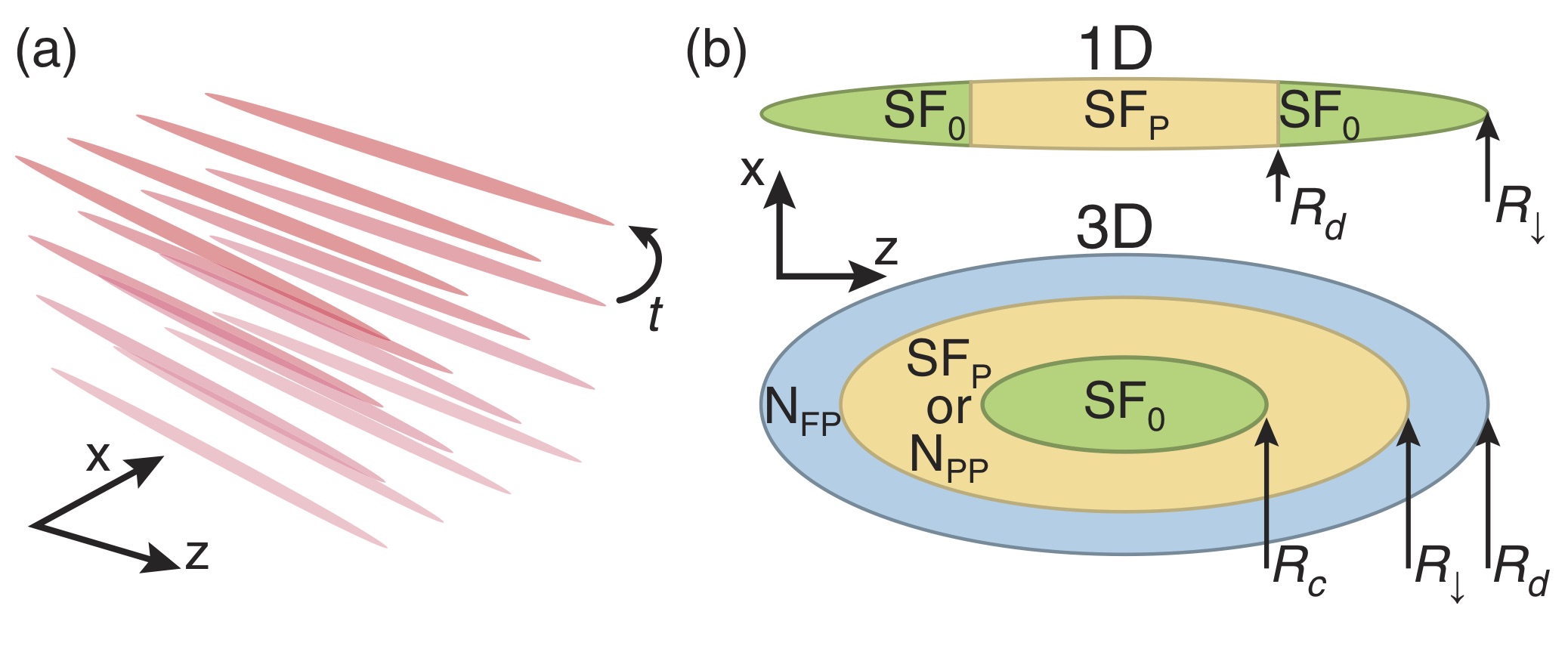}
\caption{(a) An array of 1D tubes formed by a 2D optical lattice. By decreasing the optical lattice depth, the intertube tunneling rate $t$ is increased. In this way the system can be gradually tuned from a quasi-1D to a quasi-3D regime. (b) Phase separation in a trapped Fermi gas in quasi-1D (top) and quasi-3D (bottom) regimes, at zero temperature and a small spin imbalance. The phases are: $\mathrm{SF_P}$ -- FFLO superfluid, $\mathrm{SF_0}$ -- standard BCS superfluid, $\mathrm{N_{PP}}$ -- an unpaired phase with spin imbalance, $\mathrm{N_{FP}}$ -- an unpaired normal phase. Arrows indicate phase boundaries at the different radii $R$. Reproduced with permission from  \cite{2016RevellePRL}. Copyright 2016, American Physical Society. }
 \label{Fig5} 
\end{figure}

Recently the 1D-3D crossover scenario was successfully realized experimentally with $^6$Li atoms confined in an array of 1D tubes \cite{2016RevellePRL}. The array of 1D traps was formed with a 2D optical lattice and the transverse tunneling rate could be tuned by changing the 2D lattice depth (Fig.~\ref{Fig5}a). The quasi-1D and quasi-3D regimes could be distinguished by the local spin polarization at the midpoint of the central 1D tube (Fig.~\ref{Fig5}b): a partially polarized core corresponded to the quasi-1D regime, and an unpolarized core indicated a quasi-3D regime. It was found that the critical tunneling value, corresponding to the transition between the quasi-1D and quasi-3D regime, was approximately $t_c \approx 0.025\epsilon_b$ (where $\epsilon_b$ is the pair binding energy). 

Finally, it is worth noting an alternative approach to the 1D-3D crossover was proposed in \cite{2016DuttaPRA}. Here, only a single 1D tube is considered, and the parameter controlling the dimensionality is the chemical potential $\mu$. When $\mu$ is small enough, transverse movement is confined to the lowest oscillator level, {\it i.e.}, the usual condition for quasi-1D dimensionality is fulfilled. For large enough $\mu$ transverse modes are accessible and the dynamics become locally 3D. The authors find that strong interactions, which mix single-particle levels, cause 3D-like behavior to occur at all densities. 

\subsection{Mass-imbalanced mixtures}


A parallel direction of research on unconventional pairing concentrates on the relationship between mass-imbalanced 1D systems and FFLO. In the typically considered ultra-cold systems, the source of mismatch between the two spin Fermi momenta in the ultra-cold system is the imbalance between spin populations, which leads to a difference in the chemical potential $\mu$ and thus, in the magnitude of the Fermi momenta. However, an alternate way to induce the difference between Fermi momenta is by using components with different masses. The most straightforward approach is creating a mixture of different atomic species with varying masses \cite{2018KinnunenRepProgPhys}. Alternatively, one can create a system confined in spin-dependent optical lattices, where the two spin components exhibit different tunneling amplitudes and thus different "effective mass". Such an effect can be achieved, for example, by the use of a magnetic field gradient modulated in time \cite{2015JotzuPRL}. The occurrence of the FFLO phase in mass-imbalanced 1D fermionic systems was theoretically investigated in a number of past works \cite{2005CazalillaPRL,2009WangPRA,2009BatrouniEPL,2009BurovskiPRL,2010OrsoPRL,2011RouxPRA,2012RoscildeEPL}. 

Among recent works, a three-dimensional phase diagram as a function of the mass imbalance, spin imbalance and temperature was studied in \cite{2014RoscherPRA} for a many-body system of attractive free fermions, finding that FFLO-type phases occupy a large region of the parameter space. Another recent work \cite{2017ChungPRA} theoretically studied the zero-temperature phase diagram for an attractive $^6$Li-$^{40}$K mixture confined in a 1D harmonic trap. When the two mass-imbalanced atomic species are treated as distinct pseudospin components, a greater richness of phases emerges: one can now distinguish between ``light FFLO'' or ``heavy FFLO'' phases, depending on whether the heavier or the lighter species is an excess species in the partially polarized phase. 

\subsection{Dynamical response technique}

Finally, we note several recent works that focus on the dynamical properties of the system. Although most theoretical work focuses on the ground-state properties of the system, a significant area of research focuses on the time evolution of dynamical systems after the sudden change (quench) of some parameter, such as the external potential. For instance, an FFLO Fermi gas with initial harmonic confinement which is suddenly switched off can be considered \cite{2011KajalaPRA}. The resulting cloud expansion dynamics shows a clear two-fluid behavior, where the cloud expansion velocity of one of the two components (consisting of unpaired majority fermions) is related to the FFLO momentum. This provides an experimental signature of FFLO pairing with nonzero center-of-mass momentum of the pairs. 

Another often considered area is the post-quench dynamics after a change of interaction strength. Such a scenario was considered in \cite{2015RieggerPRA} for a 1D lattice system. In particular, after a quench from zero to attractive interactions, the post-quench state shows characteristic FFLO oscillations of the pair correlation, although with exponential decay of spatial correlations. On the other hand, after a sufficiently fast quench from attractive to repulsive interactions, the initial state's FFLO correlations can be imprinted onto repulsively bound pairs if the final interaction strength is high enough. 

A different case was considered in \cite{2016YinPRA}, which analyzed the dynamics of standard BCS and FFLO states quenched from attractive to zero interactions, and analyzed the dynamics of spin and charge correlations in the post-quench system. For a quench from an initial standard BCS state, the spin correlations eventually thermalize to those of a free Fermi gas at a temperature $kT \sim U_{ini}$, while the charge component does not. On the other hand, for a quench from the FFLO state, neither component thermalizes. 

Although experimental implementation of such schemes is yet to be achieved, they are realizable with currently available techniques. For example, cloud expansion experiments with a clear resolving between clouds of single and paired atoms have been demonstrated recently \cite{2018SchergPRL}. Therefore, this approach may in the future provide the long-sought clear experimental evidence for the FFLO state. 


\section{Spin-orbit coupling}
\label{sec-soc}

With current experimental techniques, it is possible to engineer scenarios mimicking the existence of external gauge fields \cite{2011DalibardRevModPhys,2014GoldmanRepProgPhys}. In particular, artificial generation of spin-orbit (SO) coupling, i.e., the coupling between the internal and the motional degrees of freedom of a particle, has been attracting increased interest in recent years. In condensed matter systems, SO coupling plays a crucial role in the formation of exotic, topologically nontrivial phases associated closely with the quantum Hall effect \cite{2010HasanRevModPhys,2011QiRevModPhys,2013GalitskiNature}. Recent progress in spintronics has also contributed to the interest in SO coupling \cite{2004ZuticRevModPhys,2013GalitskiNature}. 

Typically SO coupling is understood as a purely relativistic effect \cite{1980ItzyksonBook,1998JacksonBook3rd} which can be explained directly from the movement of a spinful particle in the intrinsic electric field of the sample. The particle, in its comoving reference frame, experiences a magnetic field that couples to the spin. It means that the resulting spin-orbit coupling is determined by the intrinsic properties of the material and is not easily tunable. However, ultra-cold systems of neutral atoms subject to synthetic SO coupling open an alternate way to investigate this phenomenon, providing an experimentally controllable environment where the SO coupling can be precisely engineered and tuned \cite{2011LinNature,2014GoldmanRepProgPhys,2018ZhangChapter}. General reviews concerning the realization of synthetic SO coupling in ultra-cold atoms can be found in \cite{2013GalitskiNature,2015YiScienceChina,2014GoldmanRepProgPhys,2015ZhaiRepProgPhys,2019ZhangJPhysChem}.

Experimental implementation of artificial SO coupling in ultra-cold gases is already well established. The first realization of synthetic SO coupling in ultra-cold atoms came in 2011, with the realization of an SO coupling in a bosonic condensate at NIST \cite{2011LinNature}. Very soon it was also successfully engineered in 3D fermionic gases \cite{2012WangPRL,2012CheukPRL,2013WilliamsPRL,2014FuNature}. In this section, we will describe the recent experiments with synthetic SO coupling in 1D fermionic gases. 

\subsection{Experimental methods}

\begin{figure}
\includegraphics[width=0.8\linewidth]{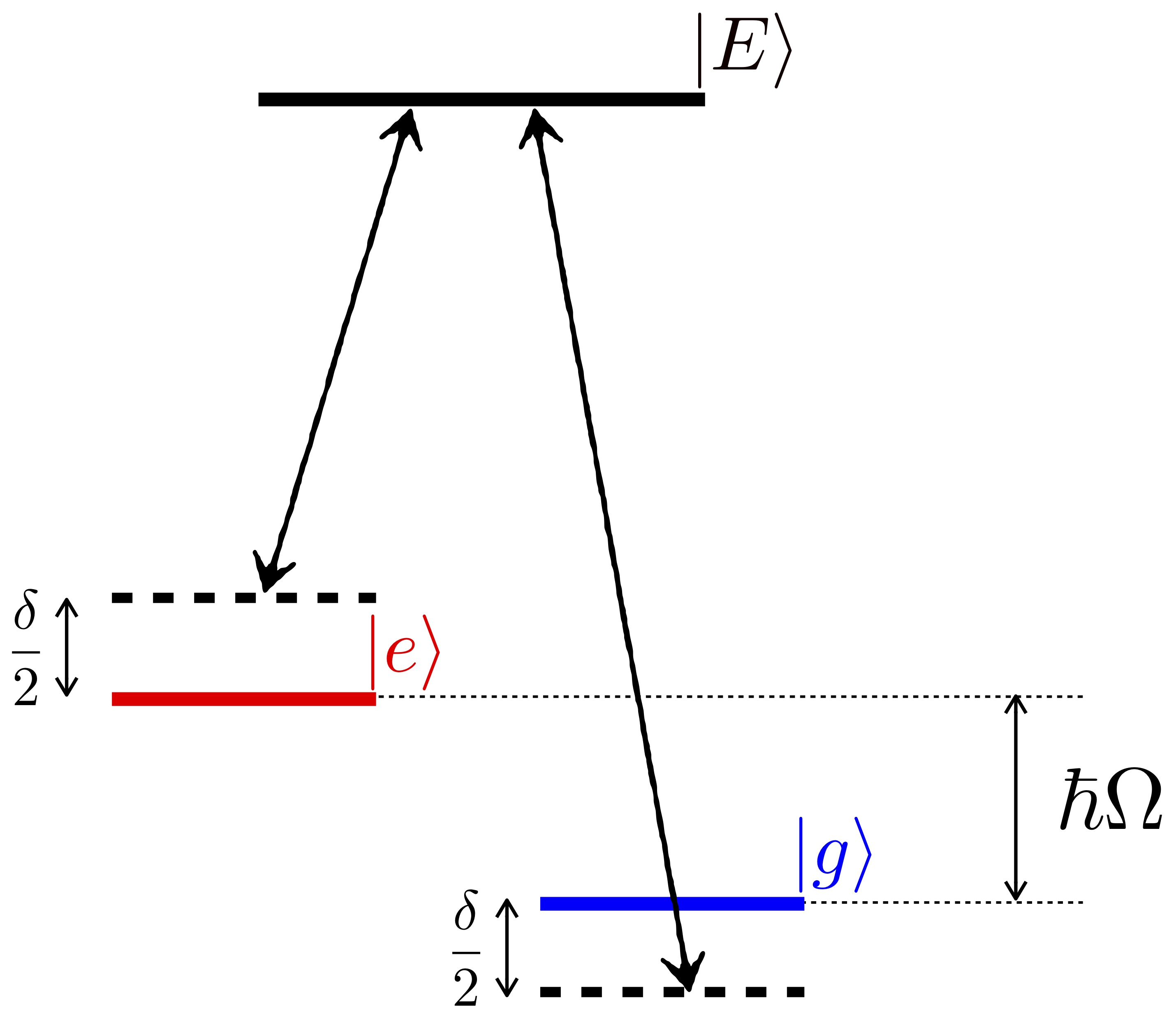}
\caption{A typical three-level Raman scheme for the generation of artificial spin-orbit coupling in ultra-cold atoms. Two internal states of the atoms, differing in energy by $\hbar \Omega$, represent the two pseudospin states $|e\rangle$ and $|g\rangle$. A pair of laser beams couples the two pseudospin states through an intermediate excited state $|E\rangle$. The lasers are detuned by $\delta$ from the Raman resonance.}
 \label{Fig6} 
\end{figure}

First, let us describe the current experimental techniques for the generation of artificial SO coupling in 1D systems. A well-established technique is the Raman laser scheme, originally proposed in \cite{2009LiuPRL} and used in the earliest experimental realizations \cite{2011LinNature,2012WangPRL,2012CheukPRL,2013WilliamsPRL,2014FuNature}. In this approach, two internal states of the ultra-cold atoms are chosen to represent pseudospin states. Then a pair of counter-propagating lasers is shined on the ultra-cold atom system, inducing a two-photon Raman coupling between the two states (see Fig.~\ref{Fig6}). Due to the conservation of momentum during the absorption and reemission of photons, the transition of an atom between the internal states is accompanied by a change in the momentum. As a result, the motion of the particle becomes coupled to the spin \cite{2013GalitskiNature}. The magnitude of the transferred momentum depends on the wavelength of the Raman beams, but it can be tuned by changing the relative angle of their intersection \cite{2014GoldmanRepProgPhys}. 

\begin{figure}
\includegraphics[width=1\linewidth]{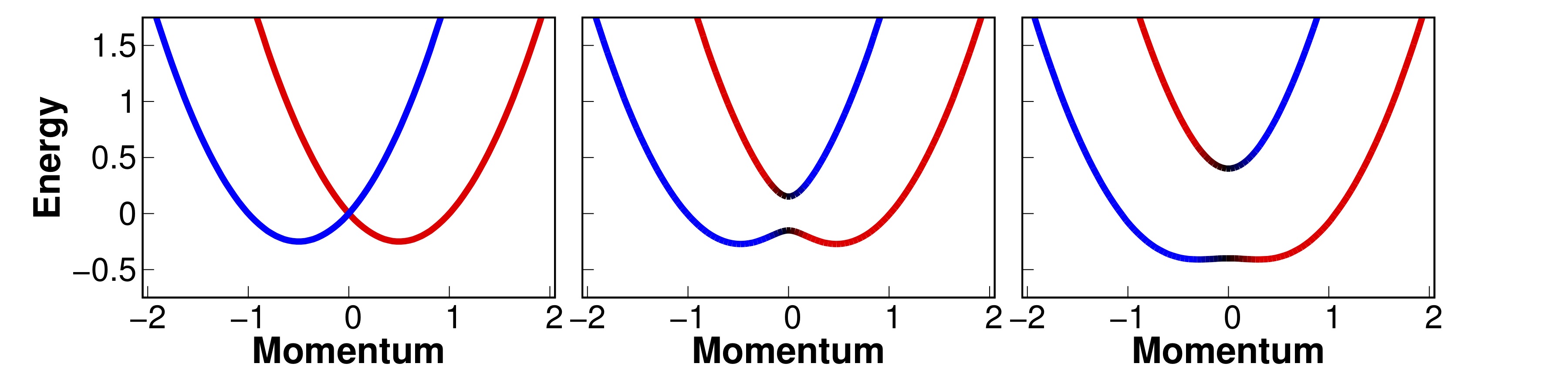}
\includegraphics[width=1\linewidth]{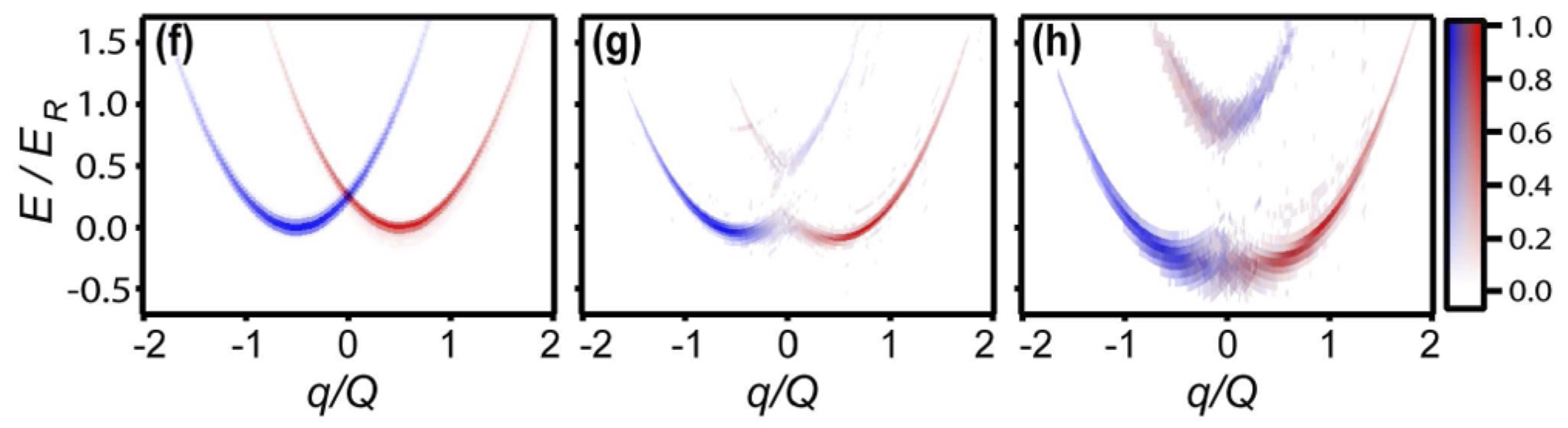}
\caption{Changes of the energy-momentum dispersion of a homogeneous ultra-cold Fermi gas in the presence of the SO coupling. Upper and bottom panels correspond to theoretical predictions and experimental results from \cite{2012CheukPRL}, respectively. With no SO coupling, the spectrum consists of two degenerate, parabolic energy bands corresponding to the two spin states. The off-diagonal coupling term (linear in momentum) causes a shift (leftmost plots), which together with the Zeeman splitting terms leads to coupling between the two energy bands and opening a gap in the spectrum (middle and rightmost plots). Colors indicate the spin composition of the states. Experimentally measured dispersions (for various Raman couplings $\Omega$) reproduced with permission from \cite{2012CheukPRL}. Copyright 2012, American Physical Society. 
}
 \label{Fig7} 
\end{figure}

This laser coupling scheme results in the realization of a one-dimensional SO coupling, equivalent to an additional term in the single-particle Hamiltonian of the general form $\propto q \hat{\sigma}_y$. Here $q$ is the momentum of the atom along the SO coupling direction, and $\hat{\sigma}_y$ is the spin Pauli matrix. Additionally, effective Zeeman terms appear in the Hamiltonian, which can be written in the general form $(\Omega/2)\hat{\sigma}_z + (\delta/2)\hat{\sigma}_y$. They are parametrized by the Raman coupling $\Omega$ and the two-photon detuning $\delta$ from the bare transition frequency (for details see \cite{2011LinNature,2014GoldmanRepProgPhys}). 

The SO coupling has a characteristic effect on the energy-momentum dispersion relation. First, due to the counter-effect of momentum transfer for opposite spins, the two bands are split and relatively shifted. Secondly, the Zeeman splitting term causes a characteristic split around zero momentum and opens a gap in the spectrum \cite{2015YiScienceChina,2018ZhangChapter}. Importantly, the resulting characteristic dispersion and spin texture of the spectrum can be probed experimentally, for example by means of a spectroscopic spin-injection technique \cite{2012CheukPRL} (see Fig.~\ref{Fig7}). 

\begin{figure}
\includegraphics[width=1\linewidth]{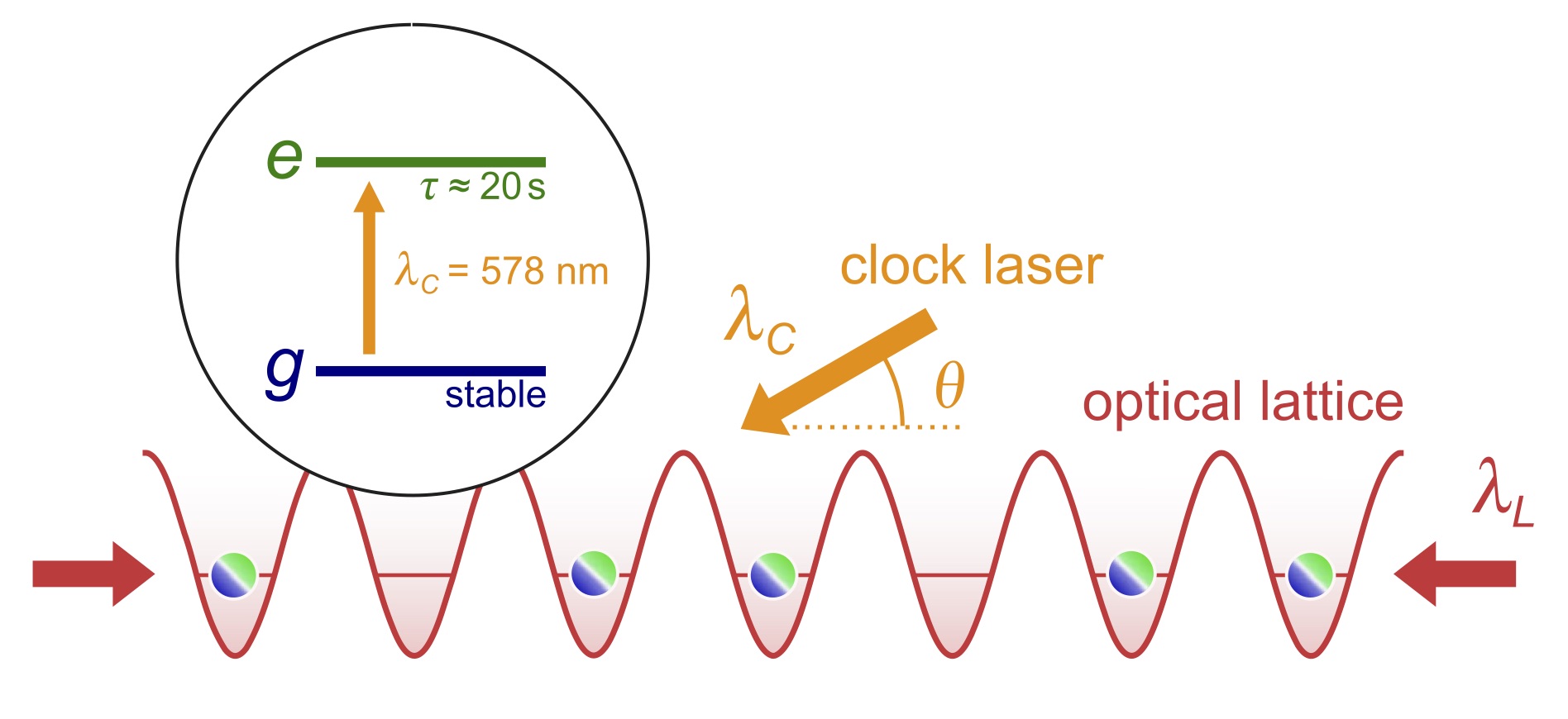}
\caption{Clock transition SO coupling setup in a 1D optical lattice, as shown in \cite{2016LiviPRL}. The fermionic $^{173}\mathrm{Yb}$ atoms are confined in an optical lattice with wavelength $\lambda_L$. The clock laser with wavelength $\lambda_C$, applied at angle $\theta$ to the optical lattice axis, drives a single-photon transition between the two states $|g\rangle ={}^1\mathrm{S}_0$ and $|e\rangle ={}^3\mathrm{P}_0$ (treated as pseudospin states). The momentum transfer $\delta_k = 2 \pi \cos(\theta)/\lambda_C$ causes a coupling between the momentum and the two pseudo-spin states. Reproduced with permission from \cite{2016LiviPRL}. Copyright 2016, American Physical Society.}
 \label{Fig8} 
\end{figure}

For atoms confined in a 1D lattice, an alternative technique of synthesizing SO coupling has been developed in recent years. In this approach, the atoms are subjected to an optical clock laser, which induces a single-photon coupling between the ground atomic state and a long-lived, metastable excited state. When the trapping lattice is set to an appropriately selected ``magic wavelength'', such that the trapping is identical for both these pseudospin states, a SO coupling emerges (Fig.~\ref{Fig8}). The SO coupling results from the fact that when the laser drives a transition between the ground and excited state, it imprints on the atom wave function an additional site-dependent phase, exactly as for an atom in an external magnetic field. Compared to the Raman technique, the advantage of this method is its simpler configuration (only one laser beam). It also avoids the detrimental effect of near-resonant intermediate states that would otherwise induce strong heating and hinder the observation of many-body effects. It should be mentioned, however, that this method is only applicable to atoms that have a necessary long-lived excited state, such as alkaline-earth atoms. Furthermore, the excited state population is vulnerable to losses due to inelastic collisions between the atoms, which may be detrimental at longer timescales \cite{2016WallPRL}. 

\begin{figure}
\includegraphics[width=1\linewidth]{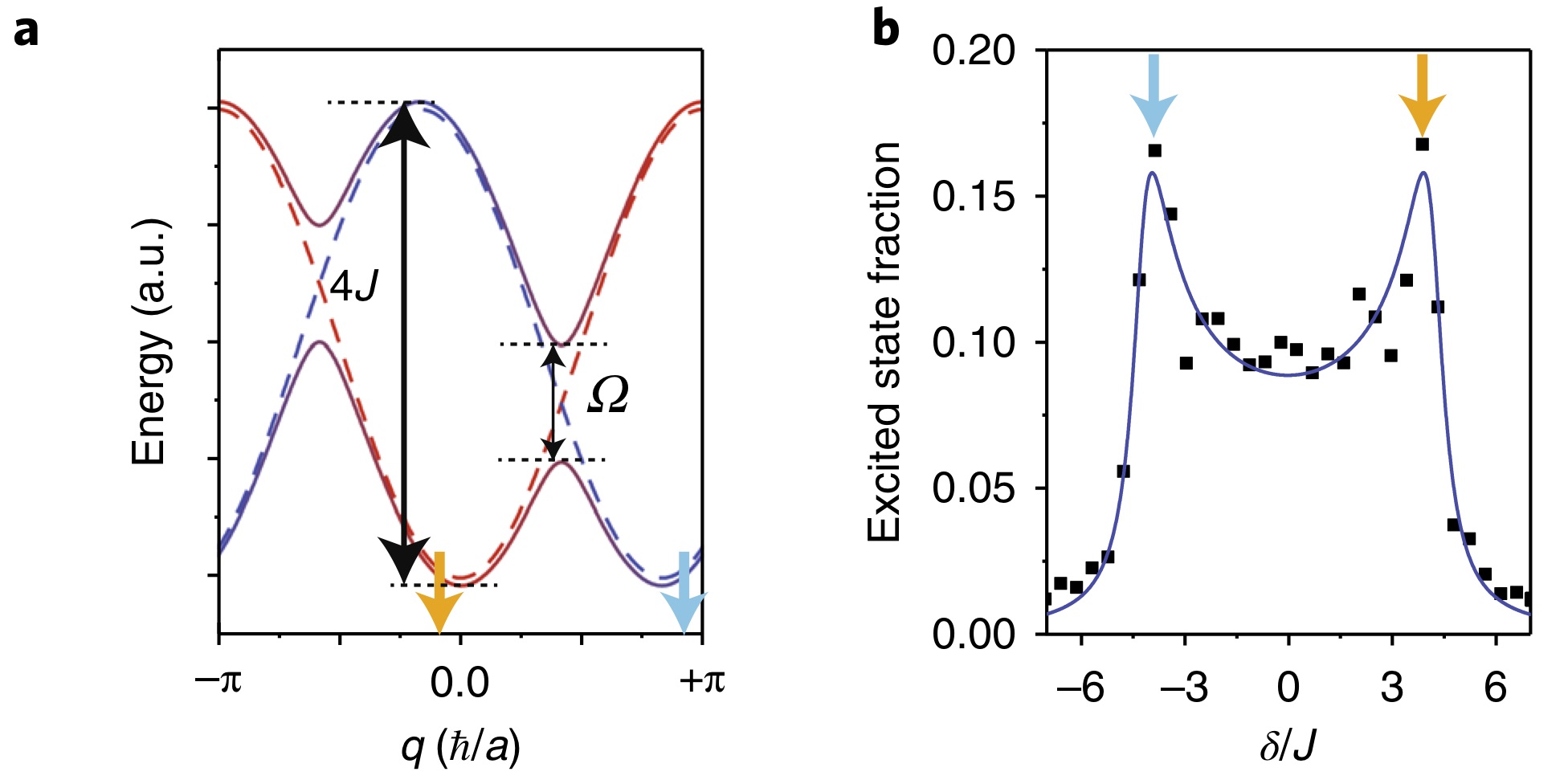}
\caption{(a) The momentum-energy dispersion spectrum of an ultra-cold Fermi gas in a lattice under SO coupling. Similarly to the free gas (Fig.~\ref{Fig7}), the bands corresponding to the two spin states are shifted and coupled. The band splitting is given by the Rabi frequency $\Omega$ of the clock transition, and the bandwidth is equal to $4J$ where $J$ is the lattice tunneling rate. The Van Hove singularities in the density of states occur at quasimomenta $q \sim 0$ and $q \sim \pi$ (indicated by the yellow and blue arrows) where a saddle point occurs in the energy difference between the two bands. (b) The $|g\rangle \rightarrow |e\rangle$ excitation spectrum of the clock transition, as a function of the detuning $\delta$ from the bare atomic transition (in units of $J$). The Van Hove singularities are manifested as peaks at the values $\delta \sim \pm 4J$ (yellow and blue arrows). Reproduced with permission from \cite{2018BromleyNatPhys}. Copyright 2018, Nature.}
 \label{Fig9} 
\end{figure}

Much like in the case of the Raman laser scheme, the quasimomentum-energy dispersion undergoes a characteristic modification in the spin-orbit coupled 1D lattice system. It can be regarded as two bands, shifted with respect to each other and coupled (Fig.~\ref{Fig9}a). This SO-coupled spectrum is characterized by divergences (Van Hove singularities \cite{1953VanHovePR}) in the density of states, which, as explained in \cite{2017KolkowitzNature}, occur at saddle points in the energy difference between the band dispersion curves. This results in the appearance of characteristic peaks in the excitation spectrum, at detunings comparable to the bandwidth (Fig.~\ref{Fig9}b). They can be used as a spectroscopic signature of the SO coupling \cite{2016WallPRL,2016LiviPRL,2017KolkowitzNature,2018BromleyNatPhys}. 

Apart from the two techniques described above, other methods have been proposed for SO generation. One proposal involves generating the effective SO coupling by periodic spin-dependent driving of atoms trapped in a lattice via a time-dependent magnetic field. In this way, the atom tunneling amplitudes become spin-dependent, and as a result, the characteristic SO splitting of the energy spectrum appears. The strength of the resulting SO coupling can be tuned by adjusting the driving amplitude \cite{2014StruckPRA}. 

Another example is the so-called Raman lattice scheme proposed in \cite{2013LiuPRL} and later implemented experimentally in \cite{2018SongSciAdv}. In this approach, two laser beams are used. One laser beam generates an optical lattice. The other perpendicular beam overlays the lattice with a periodic Raman potential inducing spin-flipping hopping between the lattice sites. It thus leads to effective spin-orbit coupling. In this approach, both beams can be generated by a single laser source, which simplifies the experimental setup. 

\subsection{Experimental realizations}

We now proceed to describe the recent experimental achievements of spin-orbit coupled 1D Fermi gases. We start by listing the recent successful implementations of the clock lattice technique for generating the SO coupling. In an experiment by the Fallani group in LENS \cite{2016LiviPRL}, a gas of ultra-cold $^{173}$Yb atoms was confined in a 1D magic wavelength lattice potential, with identical band structures for both internal states $|g\rangle = {}^1\mathrm{S}_0$ and $|e\rangle = {}^3\mathrm{P}_0$ chosen as the spin states. A clock laser along the lattice direction generated coherent coupling between the $|g\rangle$ and $|e\rangle$ states (Fig.~\ref{Fig8}). In this case, the clock laser transition was used both to implement the SO coupling and to probe the system spectroscopically. In particular, the authors confirmed that -- with increasing SO coupling strength -- the excitation spectrum of the clock transition displays a pair of characteristic peaks, corresponding to the Van Hove singularities. 

A similar experiment with SO-coupled 1D Fermi gas has been performed with $^{87}$Sr atoms in JILA \cite{2017KolkowitzNature}. In particular, the authors demonstrated that it is possible to selectively prepare atoms with particular quasimomenta $q$, thanks to the $q$-dependence of the clock transition frequency. In another recent experiment in JILA \cite{2018BromleyNatPhys} the authors focused on the effects of strong many-body interactions in the SO-coupled system, analyzing the influence of the interactions on the collective spin dynamics. 

\subsection{Artificial dimensions}

\begin{figure}
\includegraphics[width=1\linewidth]{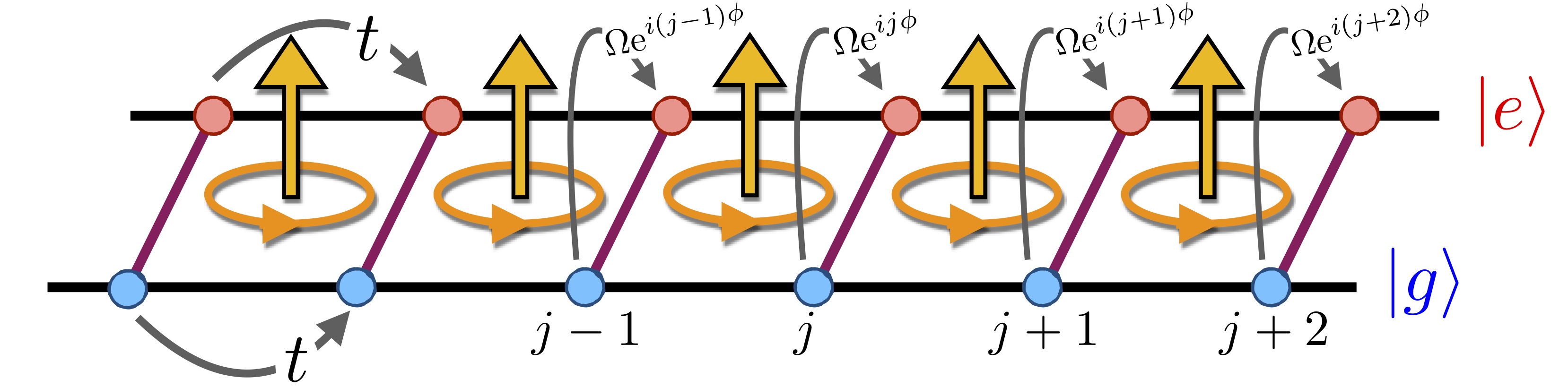}
\caption{A synthetic two-leg ladder structure, realized by atoms in a 1D lattice which have two pseudospin states $|g\rangle$ and $|e\rangle$ coupled by a clock laser transition. Hopping between the sites of the lattice takes place with a real amplitude $t$. The coupling between states $|g\rangle$ and $|e\rangle$ is equivalent to hopping along a synthetic spin dimension. Due to the artificial spin-orbit coupling, this interleg hopping has a complex amplitude with magnitude $\Omega$ and phase $j\phi$, which differs depending on lattice site index $j$. The yellow arrows indicate the effective artificial magnetic flux $\phi$ that pierces each plaquette of the ladder. Adapted with permission from \cite{2016WallPRL}. Copyright 2016, American Physical Society.}
 \label{Fig10} 
\end{figure}

An interesting aspect of such one-dimensional lattice experiments is that the spin degree of freedom can be interpreted as a ``synthetic dimension'', and transitions between the spin states can be interpreted as hoppings along this dimension \cite{2012BoadaPRL,2014CeliPRL,2015PricePRL}. In this framework, a 1D lattice loaded with fermions of ${\cal N}$ spin components is interpreted as a 2D ``ladder'' with ${\cal N}$ ``legs'' (Fig.~\ref{Fig10}). If the atoms are subject to an artificial spin-orbit coupling, the hopping in this synthetic dimension becomes complex, with a phase that depends on the lattice site index. The phase imprinted by the spin coupling varies between neighboring sites, with a value dependent on details of the 1D lattice potential and the artificial SO coupling gauge field. Then, in the synthetic dimension picture, the SO coupling corresponds to an effective magnetic field flux piercing each plaquette of the ladder \cite{2016WallPRL}. The Hamiltonian of such a system (with ${\cal N} = 2$ legs) can be written as the Harper-Hofstadter ladder Hamiltonian of the form \cite{2016LiviPRL}
\begin{multline} 
 H = \sum_j \left[\sum_\alpha  -t\,\hat{c}^\dagger_{j,\alpha} ( \hat{c}_{j-1,\alpha} + \hat{c}_{j+1,\alpha} ) \right. \\ \left. - T_j\hat{c}^\dagger_{j,e} \hat{c}_{j,g} -T_j^* \hat{c}^\dagger_{j,g} \hat{c}_{j,e}\right]
\end{multline}
where $t$ is the tunneling amplitude between neighboring sites on the same leg, $T_j = -\Omega e^{-ij\phi}$ is the site-dependent tunneling amplitude between two different legs, and the operator $\hat{c}_{j,\alpha}$ annihilates a fermion on site $j$ on leg $\alpha \in\{ e,g\}$. Frequency $\Omega$ is related to the Rabi frequency associated with the clock excitation, and $\phi$ is the effective magnetic flux per plaquette.

Such synthetic ladders with accompanying SO coupling have attracted interest due to their potential application to study topologically nontrivial states of matter in ultra-cold atoms that are not attainable in standard condensed-matter systems. For example, the two-leg ladder offers a means to realize the Creutz ladder model, one of the most important minimal models that can realize topological insulator phases \cite{1999CreutzPRL,2017JunemannPRX}. Detailed reviews on the realization of topological phases with ultra-cold atoms can be found in \cite{2016GoldmanNatPhys,2019CooperRevModPhys,2018ZhangAdvPhys}.

However, an even more interesting possibility offered by the synthetic dimension framework is using 1D lattices to emulate 2D systems \cite{2014CeliPRL}. The synthetic ladder can be interpreted as a fragment (a strip) of a larger 2D lattice. With the addition of the effective magnetic flux from SO coupling, the ladder system can emulate the physics of the topologically nontrivial Harper-Hofstadter lattice model \cite{1976HofstadterPRB}, which describes charged particles in a 2D lattice in a uniform magnetic field. In fact, it can be shown theoretically that a two-leg ladder with SO coupling can accurately reproduce the energies and wave functions of the edge states of a real Hofstadter lattice \cite{2014HugelPRA}. 

\begin{figure}
\includegraphics[width=1\linewidth]{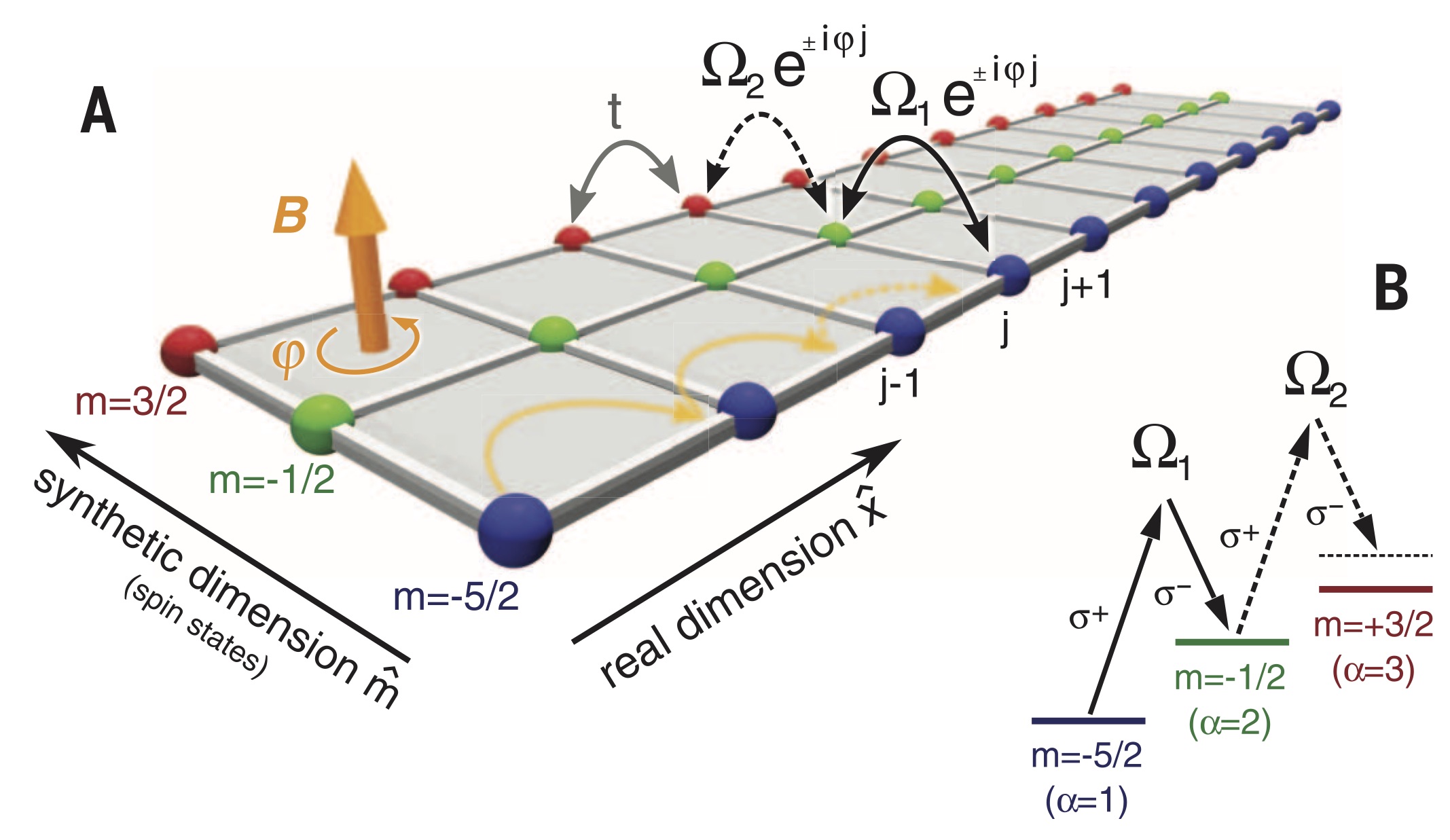}
\caption{The realization of a Harper-Hofstadter strip by means of a 1D lattice with a synthetic spin dimension. See the main text for details. Reproduced with permission from \cite{2015ManciniScience}. Copyright 2015, Science.}
 \label{Fig11} 
\end{figure}

\begin{figure}
\includegraphics[width=\linewidth]{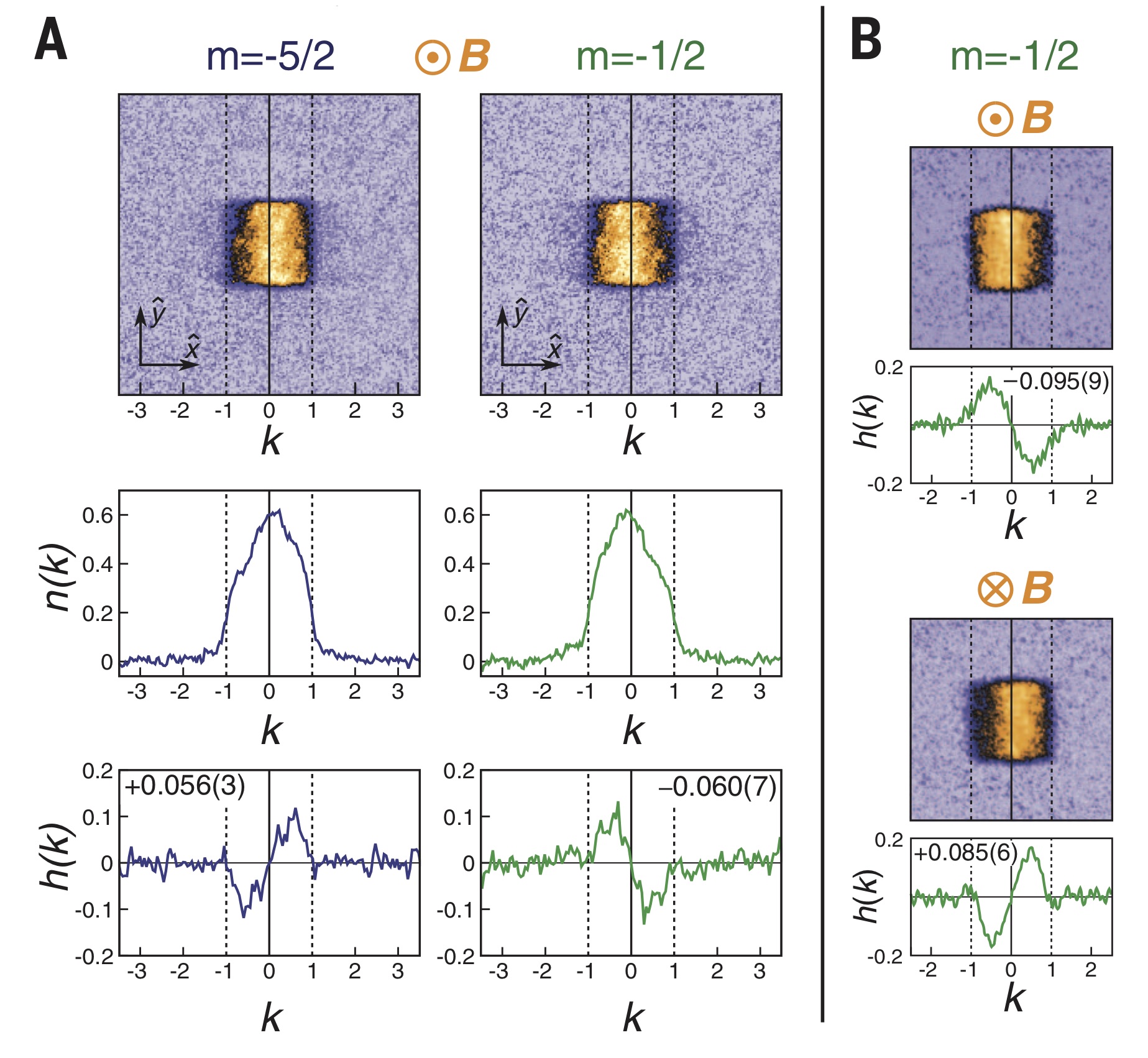}
\includegraphics[width=\linewidth]{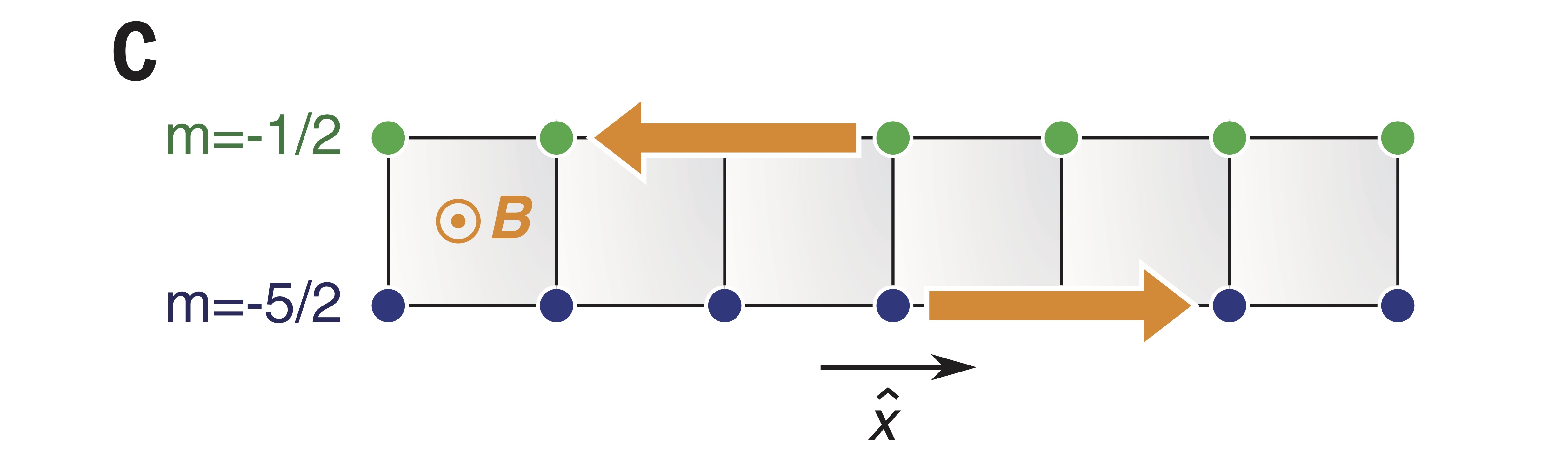}
\caption{Uncovering chiral edge currents of fermionic atoms in a two-leg ladder with effective magnetic flux. (a) Top: Time-of-flight images representing the momentum distribution of atoms in the two pseudospin states $m=-5/2$ and $m=-1/2$. Middle: Integrated momentum distributions $n(k)$. Bottom: The imbalances $h(k) = n(k) - n(-k)$. The nonzero imbalance reveals the presence of a chiral current for the atoms in a given pseudospin state, with the opposite directions for both pseudospins. (b) The momentum distribution and imbalance $h(k)$ for atoms in the $m = -1/2$ state, for two opposite directions of the effective magnetic field. It can be seen that the direction of the chiral current is inverted as the effective field is turned in the opposite direction. (c) Visualization of the two chiral currents (orange arrows) along the two legs of the ladder that correspond to the two spin states. Reproduced with permission from \cite{2015ManciniScience}. Copyright 2015, Science.}
 \label{Fig12} 
\end{figure}

\begin{figure}
\includegraphics[width=\linewidth]{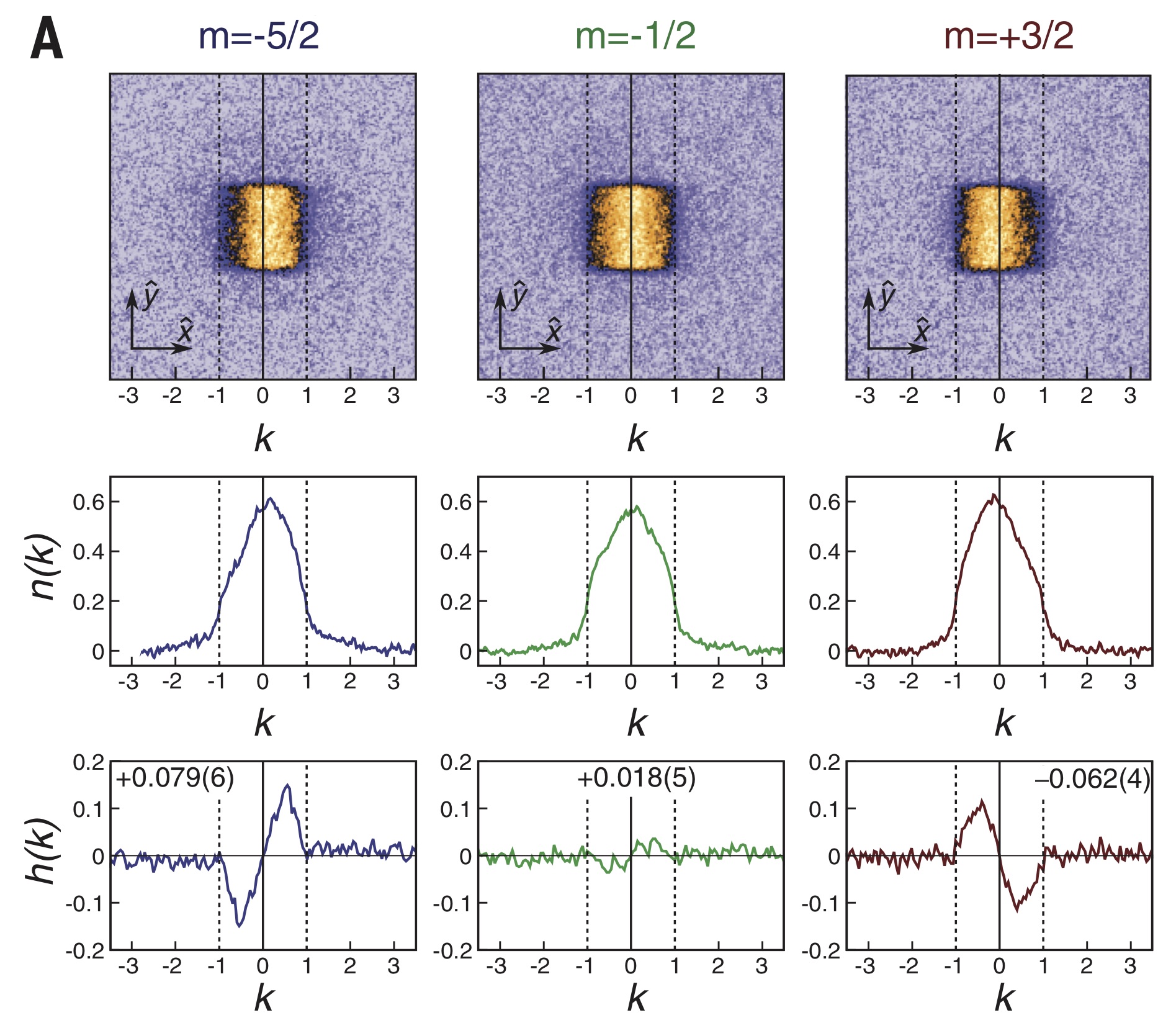}
\includegraphics[width=\linewidth]{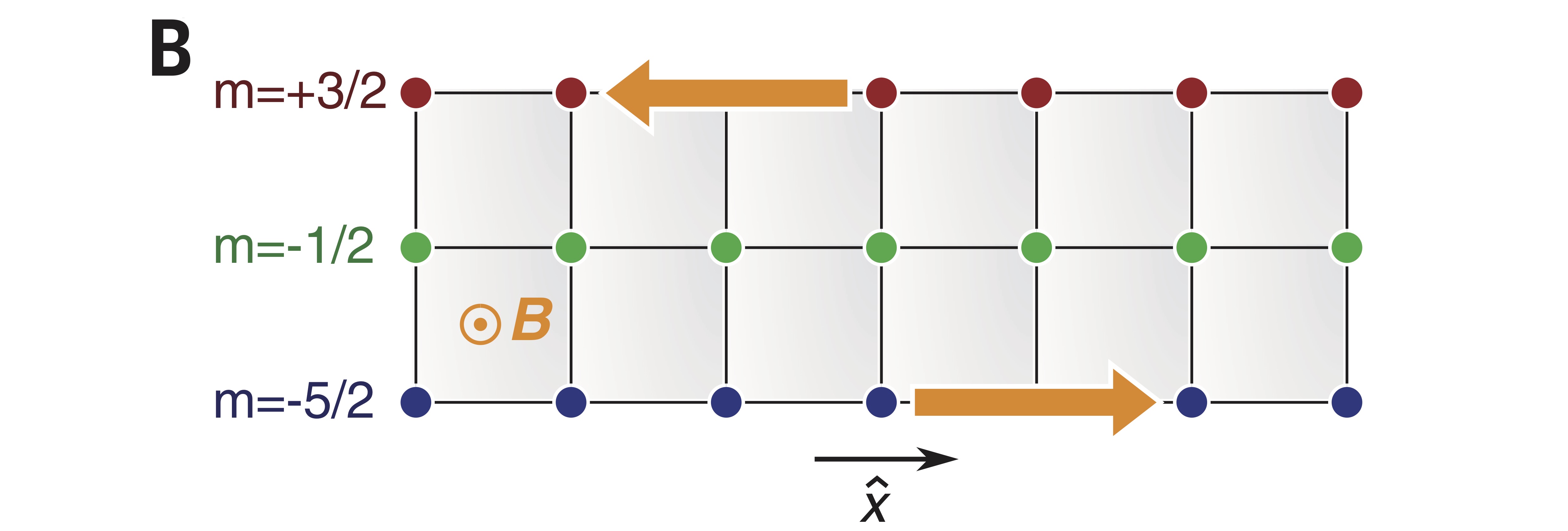}
\caption{Same as Fig.~\ref{Fig12}, but for a ladder with three legs corresponding to the pseudospin states $m=-5/2$, $m=-1/2$ and $m=+3/2$. Chiral currents are only present for atoms on the two ``edge'' legs ($m=-5/2,+3/2$), while the ``middle leg'' ($m=-1/2$) is characterized by a net zero current. Reproduced with permission from \cite{2015ManciniScience}. Copyright 2015, Science.}
 \label{Fig13} 
\end{figure}

An excellent demonstration of these possibilities is given by the experiment by the Fallani group in LENS \cite{2015ManciniScience}. Using $^{173}$Yb atoms in a 1D lattice, with Raman laser coupling between two or three distinct spin states, the authors realized a two- or three-leg ladder geometry (see Fig.~\ref{Fig11} for a pictorial view). For the two-leg ladder case, spin-resolved measurement of momentum distributions revealed the presence of edge chiral currents, travelling in opposite directions along the two legs (Fig.~\ref{Fig12}). These currents, which can be detected by analyzing spin-resolved momentum distributions, are analogous to the topological chiral modes running along the edge of the 2D Hofstadter lattice. In fact, it can be easier to experimentally investigate such edge-localized phenomena in such a 1D simulator as opposed to a real 2D structure, since the momentum distribution for each spin component can be measured individually \cite{2019CooperRevModPhys}. By increasing the number of coupled spin states from two to three, one obtains a three-leg ladder geometry, which is an even closer approximation of a strip of a 2D system. Compared to a two-leg ladder, which is ``all edge and no bulk'', the three-leg ladder has a ``bulk'' in the form of the ``central'' leg. Momentum distribution measurements reveal that no net chiral current is present in this ``bulk'' leg (Fig.~\ref{Fig13}). The experiment serves as a remarkable demonstration of how the physics in a 1D spin-orbit coupled lattice system can be mapped onto those of a 2D system. 

\begin{figure}
\includegraphics[width=1\linewidth]{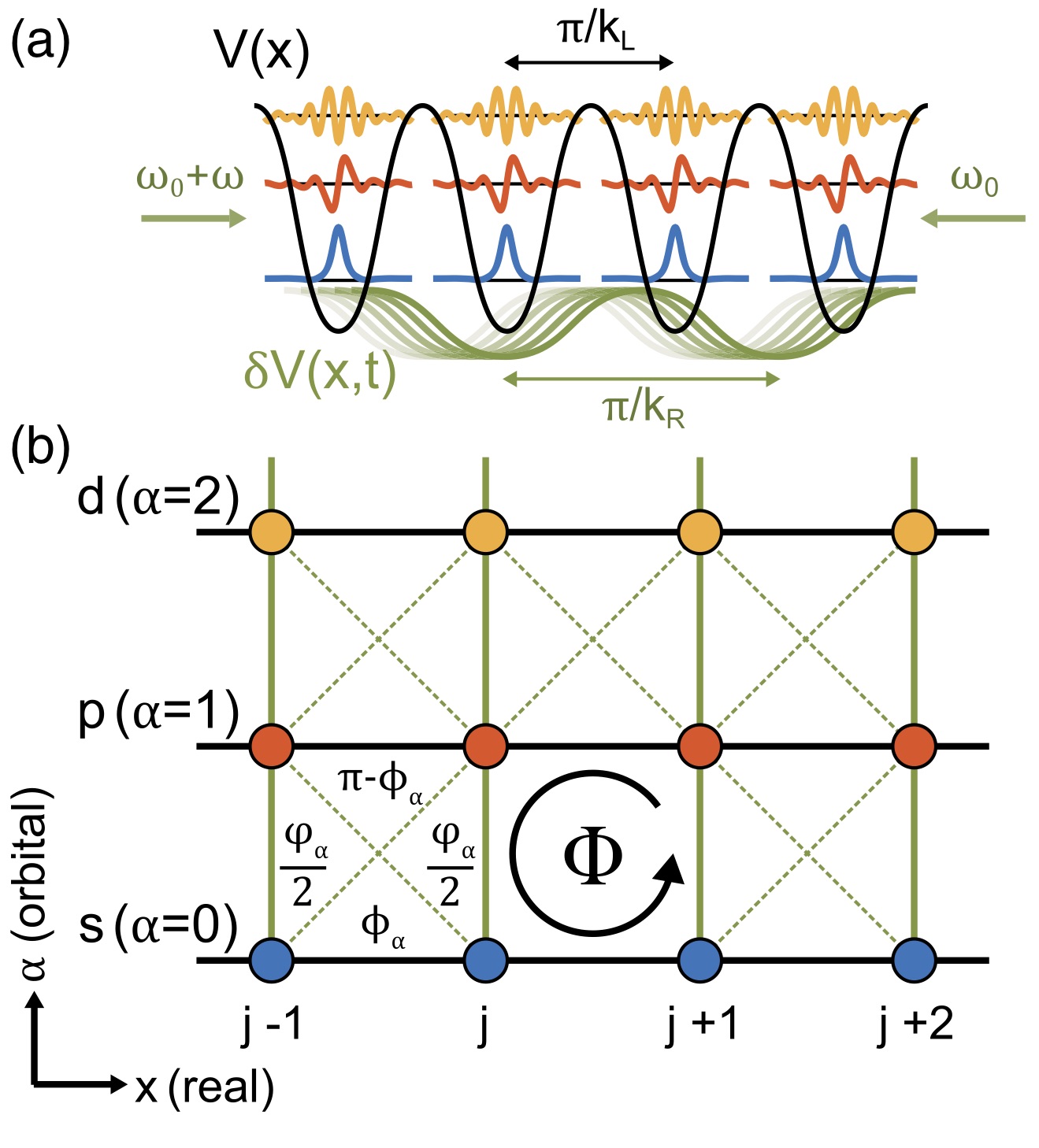}
\caption{(a) Schematic of the experimental setup realizing orbital-momentum coupling in a 1D optical lattice. The stationary 1D lattice potential $V(x)$ with lattice constant $\pi/k_L$ is overlaid with an oscillating lattice potential $\delta V(x,t)$ with lattice constant $\pi/k_R$. The moving lattice induces two-photon Raman transitions between the different orbital states of the lattice. (b) The lattice as a ladder, with the orbital states $s,p,d$ playing the role of the synthetic dimension. The particles can hop along the real dimension $x$ (black solid lines) and along the orbital dimension $\alpha$ (green solid lines), but ``diagonal'' hopping is possible as well (dashed lines). Because of the spatial modulation of the complex tunneling amplitude, an effective magnetic flux $\Phi = 2\pi(k_R/k_L + 1)$ per plaquette is created. The diagonal hopping additionally divides each plaquette into four sub-plaquettes with the magnetic flux distributed between them. Reproduced with permission from \cite{2018KangPRL}. Copyright 2018, American Physical Society.}
 \label{Fig14} 
\end{figure}

With regard to artificial ladder geometries, it is worth mentioning that a more complex ladder structure was recently achieved experimentally in Seul \cite{2018KangPRL}. In a 1D lattice with $^{173}$Yb atoms, the authors realized a three-leg \emph{cross-linked ladder}: a ladder that allows hopping between lattice sites with a simultaneous change of orbital, corresponding to diagonal hopping across the ladder plaquettes (Fig.~\ref{Fig14}b). The system was implemented by overlaying the trapping optical lattice with a periodically oscillating lattice potential, generated by a pair of Raman lasers with different frequencies. This induced couplings between the first few excited orbitals of the optical lattice sites, which played the role of pseudospin, so that an ``orbital-momentum'' coupling played the role of spin-orbit coupling (Fig.~\ref{Fig14}a). The diagonal hopping was achieved by ensuring a significant overlap of the orbital wave functions corresponding to each site. 

The rapid experimental development in this area has been accompanied by theoretical developments as well. The ladder structures can also be used for more involved applications. For example, as proposed in \cite{2015MazzaNJP}, a two-leg ladder with the two legs interpreted as "particle" and "hole" states can exhibit properties similar to that of a topologically nontrivial superconducting wire. Another interesting concept has been presented in \cite{2016MugelPRA} where it was shown how a topologically nontrivial system can be implemented via a quantum walk of ultra-cold atoms on a 1D lattice. It was argued that in certain parameter regimes the system can be mapped onto the Creutz ladder \cite{1999CreutzPRL}.

\subsection{Topological superfluids}


We now move on to another possibility opened by the application of SO coupling to 1D fermions, namely, the creation of topological superfluid phases. A topological superfluid phase features Cooper pairing between the fermions (analogously to the BCS phase) but also displays nontrivial topological characteristics. In particular, it can host zero-energy edge states with properties analogous to properties of the famous Majorana fermions -- non-existing, but theoretically possible realizations of neutral particles obeying fermionic statistics being compatible with the relativistic quantum mechanics \cite{1937MajoranaNCimento,1996RyderBook2nd}. In contrast to standard Dirac particles, Majorana fermions are their own antiparticles. Although Majorana particles were never observed as quantum particles, it is commonly argued that in some specific scenarios they may give an effective and appropriate description of excitations of many-body systems. In such cases, they are particularly interesting from a quantum information perspective, as they are highly resistant to decoherence and have been suggested as a vital element in fault-tolerant quantum computation \cite{2001KitaevPhysUsp,2008NayakRevModPhys}. Topological superfluids represent a significant opportunity to generate Majorana fermions controllably. 

Majorana fermions are known to occur effectively in certain 2D superconductors characterized by $p$-wave interparticle interactions \cite{2001IvanovPRL}. A conventional 2D $s$-wave superconductor under artificial spin-orbit coupling can also harbor Majorana fermions \cite{2008ZhangPRL,2009SatoPRL,2010SatoPRB}. One-dimensional topological superfluids have been successfully created in heterostructures, consisting of a 1D semiconductor wire subject to strong spin-orbit coupling and brought in the proximity of a bulk $s$-wave superconductor \cite{2010OregPRL,2011LutchynPRL,2011FidkowskiPRB}. Experiments with such structures have uncovered evidence for the appearance of Majorana fermions in the wire \cite{2012MourikScience,2012RokhinsonNatPhys,2012DengNanoLetters,2012DasNatPhys,2013FinckPRL}.

\begin{figure}
\includegraphics[width=1\linewidth]{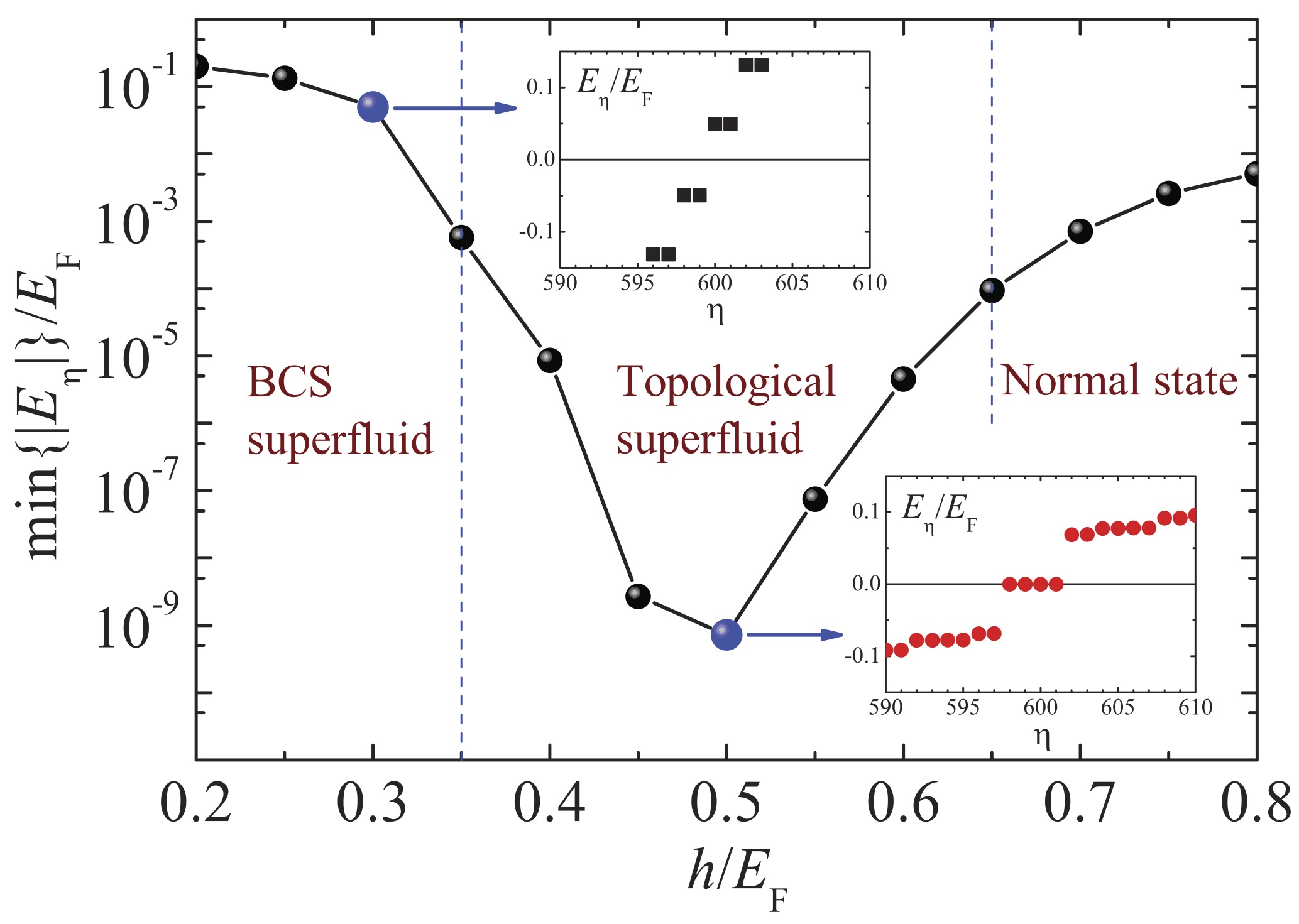}
\caption{The smallest eigenenergy $\min\{|E_\eta|\}$ of the Bogoliubov quasiparticle spectrum of a one-dimensional Fermi gas as a function of the Zeeman field $h$ under presence of the SO coupling of a fixed strength. Along with increasing Zeeman field  the system transitions from a topologically trivial BCS superfluid first to a topological superfluid, and finally to a normal state. Insets show the quasiparticle energy spectrum at $h/E_F = 0.3$ and $0.5$. It can be seen that near-zero-energy edge modes are present in the topological superfluid phase. Reproduced with permission from \cite{2012LiuPRA-Topological}. Copyright 2012, American Physical Society.}
 \label{Fig15} 
\end{figure}

Schemes for creating Majorana fermions in ultra-cold atoms have been proposed as well, for example in systems of spin-orbit coupled 1D fermions inside a background 3D BEC \cite{2011JiangPRL}. In recent years, a number of theoretical studies have explored the possibility of using SO coupling to obtain a topological superfluid phase in a solitary 1D system of attracting ultra-cold fermions, without the need to couple to external systems. In \cite{2012LiuPRA-Topological,2012WeiPRA} the case of a 1D fermionic gas in a harmonic trap, subjected to spin-orbit coupling and a Zeeman field, was analyzed. It was found that when the Zeeman field and the spin-orbit coupling are strong enough, the system can pass from a topologically trivial BCS superfluid phase into a topological superfluid phase (Fig.~\ref{Fig15}). This phase supports several zero-energy edge states, which have the Majorana-like symmetry. Analogous results were obtained in \cite{2015YangCommunTheorPhys} for 1D gas in a lattice. Additionally, it has been shown that a topological FFLO superfluid state, with a non-uniform pairing order parameter, can be obtained in this setup as well \cite{2013LiuPRA-Topological,2013ChenPRL,2017WangPRA}. 

Other works have investigated the possibility of manipulating the Majorana fermions generated in the topological superfluid. In particular, it has been proposed that Majorana fermions could be moved through the trap by manipulating the Zeeman field strength. Bound Majorana-like states can be generated at desired locations as well, by inserting impurities into the system \cite{2012LiuPRA-Manipulating,2013LiuPRA-Impurity}. Other theoretical proposals involve dark solitons in the superfluid, which can also support Majorana fermions bound to their locations. Thus, they provide an indirect way to manipulate Majorana fermions or to identify the topological nature of the state through the filling status of the solitons \cite{2014XuPRL,2015LiuPRA,2019FanPRA}. 

It is also worth noting an interesting proposal for obtaining topological superfluids which was made in \cite{2015YanSciRep}. The work considers a 1D lattice with SO coupling that realizes a ladder geometry. It is proposed that two chiral edge states on the opposite legs of the ladder can undergo Cooper pairing, leading to a BCS-like superfluid phase with zero-energy Majorana modes localized at ends of the lattice. 

Finally, we wish to draw attention to a recent experimental realization of a symmetry-protected topological (SPT) phase with fermionic $^{173}$Yb atoms in a 1D lattice in \cite{2018SongSciAdv}. SPT phases are a subset of topological phases, distinguished by the fact that, while ordinary topologically ordered phases are robust against any local perturbations, SPT phases remain intact only against perturbations that respect specific protecting symmetries. Theoretical schemes for obtaining SPT phases in 1D Fermi systems were considered for spin-orbit coupled fermions in a Raman lattice \cite{2013LiuPRL}  as well as for 1D fermions with SO coupling induced by Raman laser couplings \cite{2017ZhouPRL}. From the experimental point of view, the Raman lattice scheme was implemented and an SPT phase was successfully created in the 1D Fermi system \cite{2018SongSciAdv}. Strikingly, when the confining lattice potential was spin-dependent, the obtained topological phase was one of a new, exotic type, outside of the traditional Altland-Zirnbauer classification \cite{1997AltlandPRB} which is typically used to classify 1D SPT phases. 
 
\section{Higher-spin systems}
\label{sec-sun}

Due to the obvious historical reasons, most research on fermionic systems concerns spin-1/2 systems with only two distinct spin states, governed by a $\mathrm{SU}(2)$ symmetry. In this way, a very close analogy to the electronic systems is kept. However, current experimental achievements in the field of atomic physics allow the exploration of higher-spin systems, which can be used to realize a rich variety of interesting phases being completely beyond the range of solid-state physics \cite{2014CazalillaRepProgPhys,2016CapponiAnnPhys,2019SowinskiRepProgPhys}. An important subset of such higher-spin systems are $\cal{N}$-component systems with an $\mathrm{SU}({\cal N})$ symmetry. The physics of higher $\mathrm{SU}({\cal N})$ symmetries are of interest to many branches of physics and can lead to new connections with high-energy physics. For instance, an $\mathrm{SU}(3)$ symmetry underlies the description of quarks in quantum chromodynamics \cite{2007GreinerBook}, while an $\mathrm{SU}(6)$ symmetry has been used to describe the flavor symmetry of spinful quarks \cite{1964SakitaPR}. In fact, it has been proposed that systems of ultracold $\mathrm{SU}(3)$ fermions may be used to simulate some aspects of quantum chromodynamics \cite{2007RappPRL,2008RappPRB}. As another example, models with $\mathrm{SU}(4)$ symmetry can be used to study electron systems with orbital degeneracy \cite{1998LiPRL,1999FrischmuthPRL,1999AzariaPRL}.

From a theoretical point of view, the study of 1D $\mathrm{SU}({\cal N})$ fermions dates back to the work by Sutherland \cite{1968SutherlandPRL}, who extended Gaudin and Yang's 1D fermionic gas model \cite{1967GaudinPLA,1967YangPRL} to an arbitrary number of spin components, $\cal N$, giving the solution in terms of ${\cal N}$ nested Bethe ansatzes. The ground-state solution of the attractive case, which has the form of a ${\cal N}$-particle bound state, was given in \cite{1970TakahashiProgTheorPhys}. Since then, 1D $\mathrm{SU}({\cal N})$ multicomponent fermionic systems have been thoroughly explored theoretically. Examples include the three-component $\mathrm{SU}(3)$ Fermi gas \cite{2008GuanPRL,2010HePRA,2012KuhnNJP}, $\mathrm{SU}(4)$ spin-3/2 fermions \cite{2008CapponiPRA,2009GuanEPL,2012SchlottmannPRB-Spin32}, and systems with even higher symmetries \cite{2005SzirmaiPRB,2007BuchtaPRB,2008LiuPRA,2008SzirmaiPRB,2010GuanPRA,2011ManmanaPRA,2011YinPRA,2012SchlottmannPRB-Arbitrary}. General reviews concerning ultra-cold fermionic systems with higher spin symmetries can be found in \cite{2014CazalillaRepProgPhys,2016CapponiAnnPhys}.

Achieving enlarged $\mathrm{SU}({\cal N})$ symmetry in condensed matter systems is usually very difficult, as it requires fine-tuning of the interaction parameters. Ultra-cold fermionic systems, thanks to their tunability, offer a very good environment for studying the higher $\mathrm{SU}({\cal N})$ physics experimentally. 

\subsection{Towards $\mathrm{SU}({\cal N})$ symmetry}

Let us look closely at the conditions necessary for obtaining $\mathrm{SU}({\cal N})$ symmetry with spin-$S$ ultra-cold atoms. A two-body interaction between two spin-$S$ fermions depends on their total spin $F$, which can assume possible values $F = 0,1,2,\ldots,2S$. Assuming contact interactions via $s$-wave collisions, the two-body interaction term between two spin-$S$ fermions can be written as \cite{1998HoPRL} 
\begin{equation}
 V(\vec{r},\vec{r'}) = \delta(\vec{r}-\vec{r'}) \sum\limits^{2S-1}_{F=0,2...} g_F P_F,
\end{equation}
where $P_F$ is a projection operator on states with total spin $F$, and the coupling constants $g_F$ depend on the $s$-wave scattering lengths $a_F$. Note that, due to fermionic quantum statistics, only even $F$ values allow for non-vanishing interactions via $s$-wave collisions. Therefore the system exhibits $S + 1/2$ distinct $s$-wave scattering lengths $a_F = a_0, a_2, ... a_{2S-1}$. The $\mathrm{SU}({\cal N})$ symmetry is obtained only when all these scattering lengths are simultaneously equal \cite{2014CazalillaRepProgPhys}. 

Alkaline-earth atoms, and atoms with a similar electronic structure such as ytterbium, are particularly well suited to this purpose \cite{2009CazalillaNJP,2010GorshkovNatPhys,2014CazalillaRepProgPhys}. In the ground state ${}^1\mathrm{S}_0$ of alkaline-earth atoms, as well as in the metastable excited state ${}^3\mathrm{P}_0$, the total electronic angular momentum is zero. As a result, the hyperfine interaction vanishes and the electronic shell configuration becomes decoupled from the nuclear spin. Since the differences in $a_F$ depend mainly on the electronic wave functions of the colliding atoms, this decoupling causes $a_F$ to become almost independent of the nuclear spin. More precisely, the nuclear-spin-dependent correction of the scattering lengths is on the order of $\sim 10^{-9}$ in the $^1\mathrm{S}_0$ state and on the order of $\sim 10^{-3}$ in the $^3\mathrm{P}_0$ state \cite{2010GorshkovNatPhys}. Thanks to this independence of scattering lengths on the spin, the system effectively exhibits a $\mathrm{SU}(2S+1)$ symmetry \cite{2016CapponiAnnPhys}. 


Alkaline-earth and alkaline-earth-like atoms have been successfully used to experimentally realize systems with higher $\mathrm{SU}({\cal N})$ symmetries. In particular, experiments in 3D and 2D setups have realized $\mathrm{SU}(6)$ symmetry with $^{173}$Yb atoms \cite{2010TaiePRL,2012TaieNatPhys,2014ScazzaNatPhys} as well as $\mathrm{SU}(10)$ symmetry with $^{87}$Sr atoms \cite{2013StellmerPRA,2014ZhangScience}. 

Of course, it should be noted that high-spin systems without a full $\mathrm{SU}({\cal N})$ symmetry can also be of significant interest. For example, a three-component fermionic system with anisotropic scattering lengths was studied theoretically in \cite{2009AzariaPRA}, and fermionic systems with spin $S \ge 3/2$ and inequal scattering lengths were investigated in \cite{2005LecheminantPRL}. In this class, a particularly interesting case is the four-component spin-3/2 system, which notably exhibits a high $\mathrm{SO}(5)$ symmetry even for inequal scattering lengths $a_0 \ne a_2$ \cite{2003WuPRL,2005WuPRL,2006WuModPhysLet}. The phases of such a spin-3/2 fermionic system (with $a_0 > 0, a_2 < 0$) were studied recently in \cite{2012BarczaPRA,2015BarczaEPL}. 

\subsection{One-dimensional realizations}
\begin{figure}
\includegraphics[width=1\linewidth]{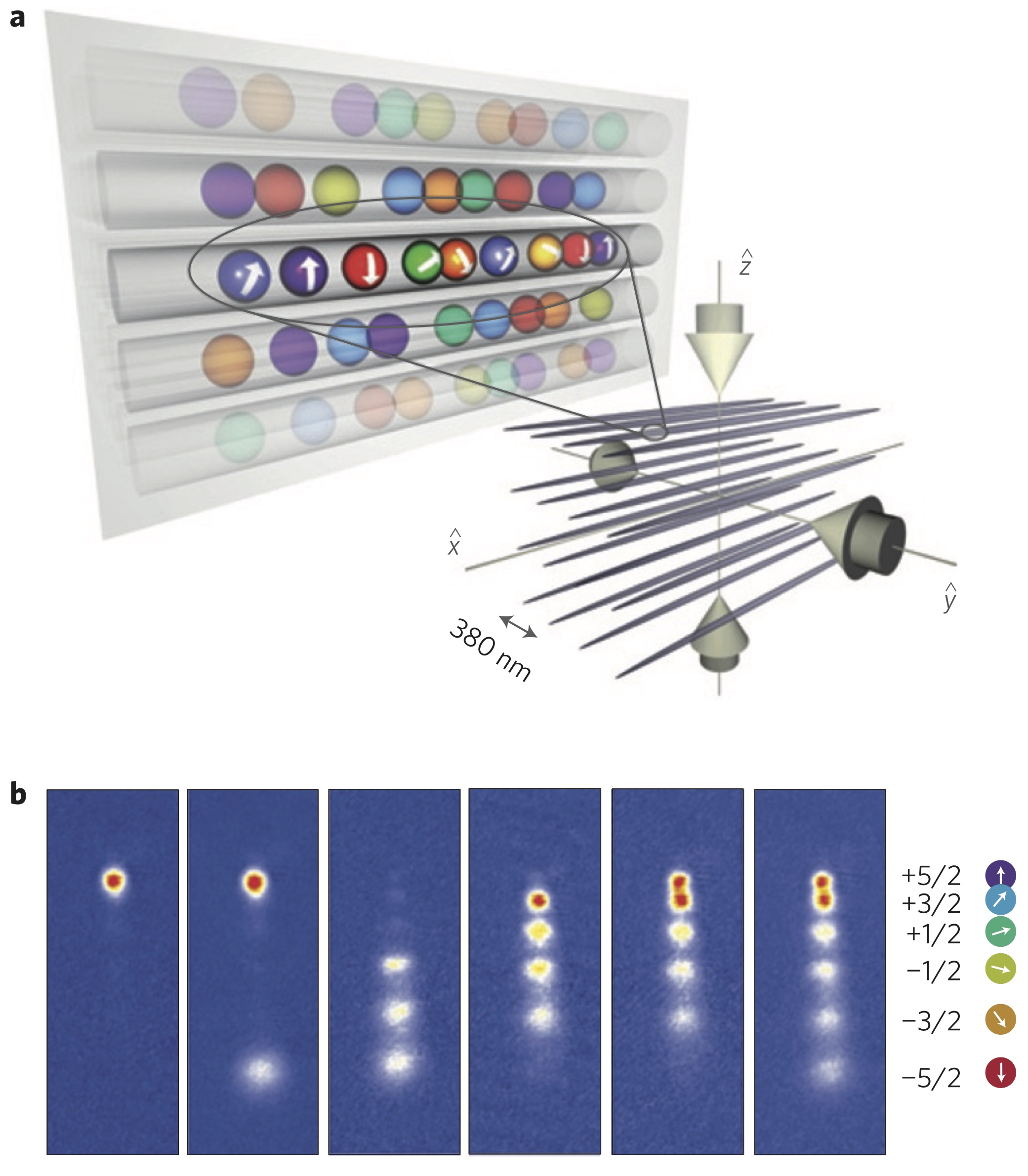}
\caption{Experimental creation of one-dimensional $\mathrm{SU}({\cal N})$ fermionic systems with tunable number of spin components. (a) A 2D optical lattice is used to create an array of independent 1D tubes of ultra-cold $^{173}$Yb atoms with up to six different nuclear spin orientations. (b) The number of spin components is fully tunable and can be determined via optical Stern-Gerlach detection. Reproduced with permission from \cite{2014PaganoNatPhys}. Copyright 2014, Nature.}
 \label{Fig16} 
\end{figure}

For 1D systems, a breakthrough experimental achievement was performed by the Fallani group \cite{2014PaganoNatPhys}. In this experiment, a one-dimensional liquid of repulsively interacting $^{173}$Yb atoms with an arbitrarily tunable number of spin components was obtained (Fig.~\ref{Fig16}a). The number of spin components $\cal N$ was set during the preparation of the sample, by means of optical spin manipulation and detection techniques. The authors have explored the physics of this system for a varying number of components, from ${\cal N} = 1$ to ${\cal N} = 6$ (Fig.~\ref{Fig16}b), while keeping the number of atoms per spin component constant. 

It was found that with an increasing number of spin components, the system properties deviate from those of a spin-1/2 Luttinger liquid that is typically used to describe 1D fermionic systems \cite{2003GiamarchiBook}. In particular, as $\cal N$ increases, the Pauli principle is increasingly less important and the system gradually takes on the properties of a system of spinless bosons (confirming earlier theoretical predictions \cite{2011YangCPL,2012GuanPRA}). This was experimentally confirmed by measuring the frequency of breathing oscillations of the cloud after a sudden change of the trap frequency. The authors have also analyzed the momentum distribution of the system and showed that it broadens monotonically as the number of components is increased. This can be explained qualitatively: as the number of spin components increases, the role of repulsions between the atoms is increased, which decreases the space available to the atoms (in a manner similar to the Pauli repulsion) and forces the occupation of higher-momentum states. The authors also probed the excitation spectra by means of the Bragg spectroscopy, finding that, for larger numbers of components, the results for the excitation frequency deviate from the predictions of the Luttinger liquid theory.

\subsection{Plethora of various phases}
\begin{figure}
\includegraphics[width=1\linewidth]{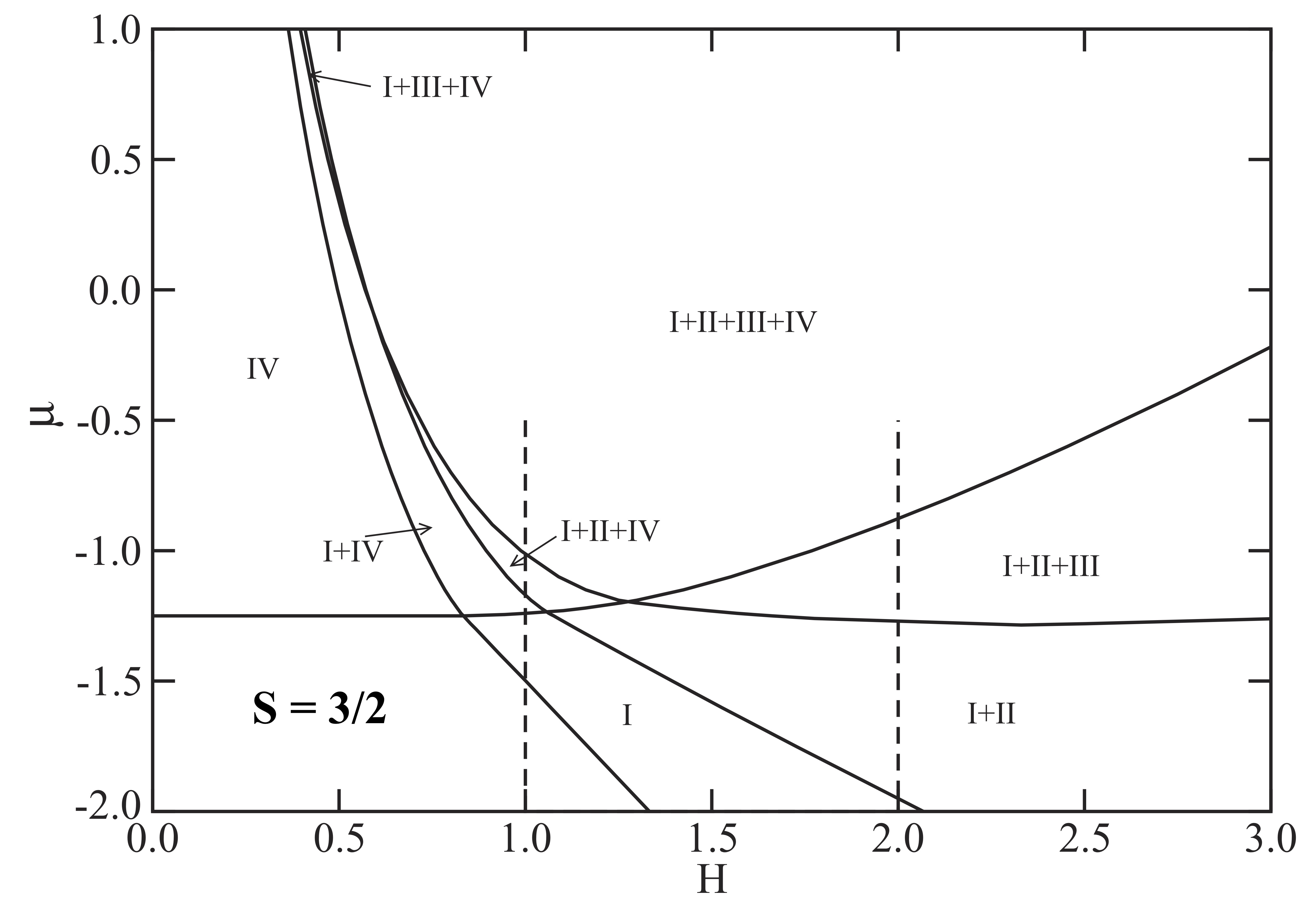}
\includegraphics[width=1\linewidth]{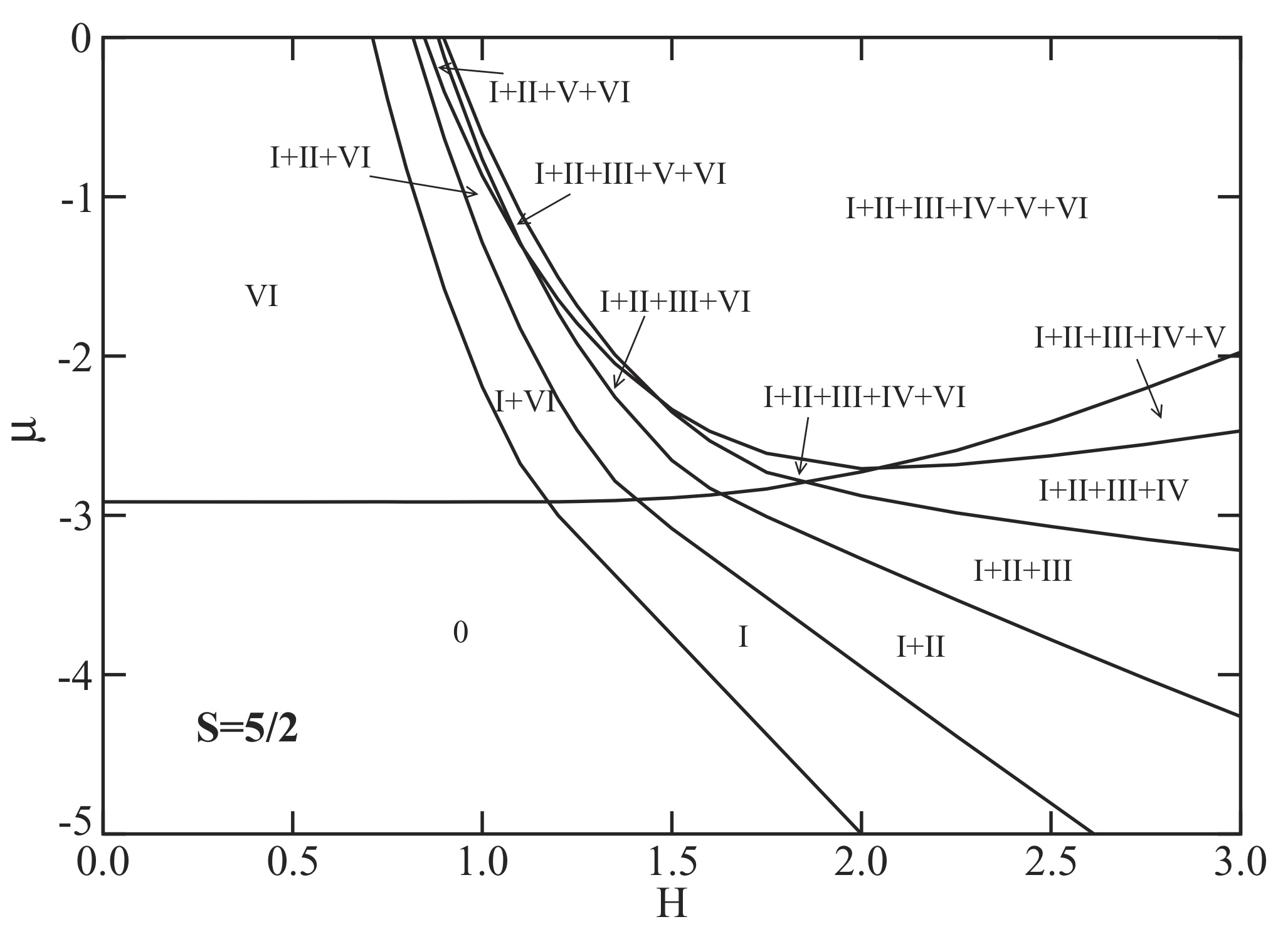}
\caption{The ground state phase diagram of a spin-3/2 $\mathrm{SU}(4)$ (upper plot) and spin-5/2 $\mathrm{SU}(6)$ (bottom plot) homogeneous one-dimensional Fermi gas with attractive interactions, in the plane of chemical potential $\mu$ vs. magnetic field $H$. The Roman numbers indicate different possible phases made up of bound states of the corresponding number of fermions. The unlabelled region is the vacuum state. Multiple roman numbers added together indicate mixed phases with coexistence of different states. The vertical dashed lines indicate the trajectories of local chemical potential for systems in a harmonic trap. For clarity, compare to the $\mathrm{SU}(2)$ phase diagram in Fig.~\ref{Fig1}. Both figures reproduced with permission from \cite{2012SchlottmannPRB-Spin32} and \cite{2012SchlottmannPRB-Arbitrary}, respectively. Copyright 2012, American Physical Society.}
 \label{Fig17} 
\end{figure}

In anticipation of future experiments, we will now point out the various interesting phases possible to create in high-spin $\mathrm{SU}({\cal N})$ systems. In the case of multicomponent systems with attractive interactions, their phase diagrams admit new types of binding beyond pair formation. Systems with ${\cal N}>2$ components allow the possibility of three-fermion (trions), four-fermion (quartets) and even larger clusters, as well as mixed phases with various combinations of such clusters \cite{2009GuanEPL}. 

To see an example of the rich possibilities, we may consider the higher-spin equivalent of the FFLO system considered in Section \ref{sec-fflo}. Let us consider a homogeneous gas with attractive interactions and a spin imbalance, subjected to a magnetic field $H$. The phase diagram for such a system with spin-1/2 was considered in the previous sections (Fig.~\ref{Fig1}). The phase diagrams for equivalent systems with higher spin symmetry are shown in Fig.~\ref{Fig17}. The spin-1/2 system admits three phases -- a polarized phase of singlet atoms, a fully paired phase, and a partially polarized phase which contains both pairs and unpaired atoms. However, a system with ${\cal N}>2$ spin components admits more phases -- a polarized phase of singlet atoms, a phase consisting of $\cal N$-fermion clusters, and numerous mixed phases in which various combinations of clusters with different particle numbers coexist. The resulting phase diagram is highly complex \cite{2012SchlottmannPRB-Spin32,2012SchlottmannPRB-Arbitrary}. 

\begin{figure}
\includegraphics[width=1\linewidth]{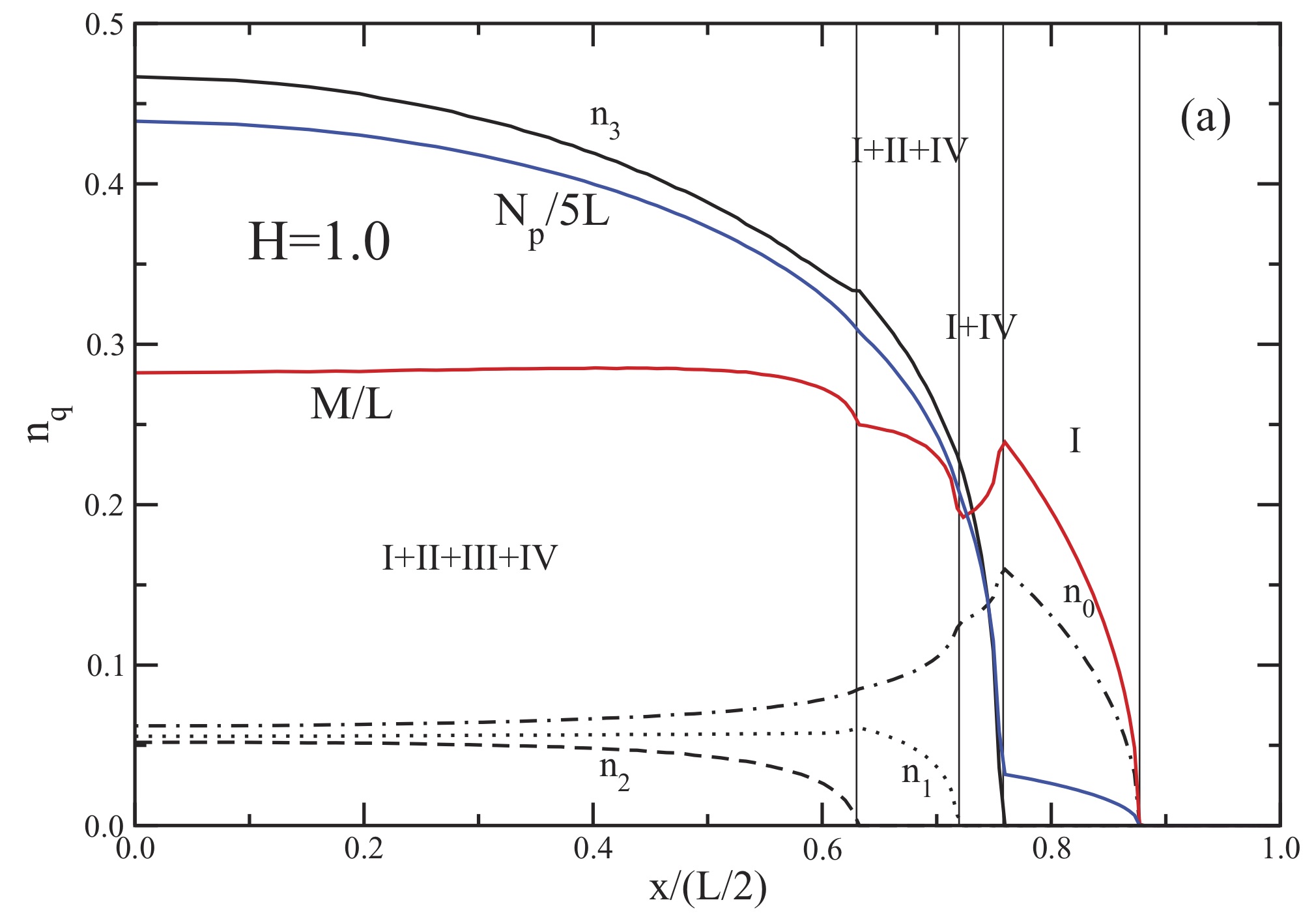}
\caption{Radial structure of a spin-3/2 $\mathrm{SU}(4)$ one-dimensional Fermi gas with attractive interactions trapped in a harmonic trap. Roman numbers indicate the phases present in the different regions of the trap. Black lines labelled with $n_q$ indicate densities of $q+1$-particle states, line labelled with $N_p/L$ is the total particle density, line labelled with $M/L$ is the magnetization density. A complex phase separation pattern is clearly visible. Reproduced with permission from \cite{2012SchlottmannPRB-Spin32}. Copyright 2012, American Physical Society.}
 \label{Fig18} 
\end{figure}

Such a complex phase diagram also allows for highly intricate phase separation in inhomogeneous systems, since a cut across the phase diagram can cross many phase boundaries \cite{2012SchlottmannPRB-Arbitrary}. Fig.~\ref{Fig18} shows an example of a complicated phase separation structure in a trapped spin-3/2 attractive gas. The phase-separated system can be described as a four-shell structure, displaying four distinct phases. It is useful to compare this case with the case of a trapped spin-1/2 attractive gas, which, as noted in Section \ref{sec-fflo}, can be described by a two-shell structure with one phase in the center and another in the wings. 

A rich variety of possible phases exists in high-spin lattice systems as well. In the tight-binding limit they can be described by the multicomponent $\mathrm{SU}({\cal N})$ variant of the Hubbard model, described by a Hamiltonian of the form \cite{2005SzirmaiPRB} 
\begin{equation}
 H = \sum_j\left[\sum_\sigma  -t\,\hat{c}^\dagger_{j,\sigma} ( \hat{c}_{j-1,\sigma} + \hat{c}_{j+1,\sigma}) + \frac{U}{2} \sum_{\sigma'\neq\sigma} \hat{n}_{j,\sigma} \hat{n}_{j,\sigma'}\right],
\end{equation}
where $\sigma,\sigma'$ are the spin components. For ${\cal N} > 2$ this model admits various interesting phases, depending on the lattice filling, the interaction strength, and the sign of the interaction (repulsive or attractive). For example, at ${\cal N} > 2$ and a filling of ${\cal N}/2$ atoms per site, the Hubbard model displays a charge-density wave phase when the interactions are attractive, and a dimerized spin-Peierls phase on the repulsive side. An extensive description of the various possible phases can be found in \cite{2016CapponiAnnPhys}. 

In the context of one-dimensional fermionic systems two aspects are particularly worth noting. One is related to the Mott transition in higher spin models. It is known that, in contrast to higher spatial dimensions, exactly in one-dimensional case the ground state of the $\mathrm{SU}(2)$ Hubbard model at half-filling is a Mott insulator for any positive interaction $U$ and it undergoes the Mott transition to the metalic phase exactly at $U=0$ \cite{1968LiebPRL,2005EsslerBook}. However, the situation is substantially different for ${\cal N} > 2$, where the transition occurs for finite repulsion $U > 0$, similarly as in higher dimensions \cite{1999AssarafPRB}. The other aspect appears in the opposite limit of high interactions $U/t \gg 1$ and exact filling of one particle per lattice site. Exactly under these conditions, the effective low-energy Hamiltonian of the system (obtained in the second-order perturbation theory) is equivalent to the $\mathrm{SU}({\cal N})$ Heisenberg model written as \cite{1999AssarafPRB,2005SzirmaiPRB,2011ManmanaPRA} 
\begin{equation}
\label{eq:heisenberg}
H = -J\sum_{i}\sum_{\mu,\nu} \hat{\mathbf{S}}_\mu^\nu(i)\hat{\mathbf{S}}_\nu^\mu(i+1),
\end{equation}
where operators $\hat{\mathbf{S}}_\mu^\nu(i)=\hat{c}_{i,\mu}^\dagger\hat{c}_{i,\nu}$ represent the $\mathrm{SU}(\cal{N})$ generators. The effective coupling $J \sim -t^2/U$ is negative and therefore it favors antiferromagnetic spin ordering. This correspondence between both systems has practical importance, as direct experimental control of the $\mathrm{SU}(\cal{N})$ antiferromagnets requires cooling to very low temperatures, $kT \lesssim J$. However, such low temperatures may be more attainable in higher-spin cold atom systems. Numerical studies indicate that, when atoms are loaded adiabatically into an optical lattice, the final temperature decreases with increasing $\cal{N}$ \cite{2012BonnesPRL,2012MessioPRL}. It comes from the fact that, in the case of atomic systems, the additional spin degrees of freedom can help to ``absorb'' the entropy from motional degrees of freedom \cite{2012TaieNatPhys} leading effectively to a lower final temperature. Higher-spin $\mathrm{SU}(\cal{N})$ systems may therefore provide an easier way to investigate exotic quantum magnetism.

Higher-spin systems also allow the possibility of extending the idea of the FFLO phase to larger fermion clusters. In the $\mathrm{SU}(4)$ system made up of an array of tubes with weak tunneling between them, a theoretical calculation shows that the mixed phases (where clusters of different length coexist) can display characteristic FFLO-like oscillations in the order parameters \cite{2012SchlottmannPRB-Spin32}. A recent theoretical investigation \cite{2017SzirmaiPRA} of the phase diagram of the $\mathrm{SU}(4)$ attractive Hubbard model at quarter filling has found two distinct FFLO-like phases, a ``paired-FFLO'' and a ``quartet-FFLO'' phase. The latter is an equivalent of the normal FFLO state, for bound particle quartets as opposed to pairs. The ``quartet FFLO'' phase appears at lower interaction strengths, but at higher interactions, it transitions into a phase-separated state where quartets and pairs coexist (see \cite{2017SzirmaiPRA} for a detailed discussion). 

\subsection{Orbital physics of higher-spin fermions}

Intriguing physics can be revealed in models which, in addition to the nuclear spin degree of freedom, explicitly take into account an additional orbital degree of freedom. For alkaline-earth atoms, the most natural candidates for this orbital degree of freedom are the electronic ground and excited states $|g\rangle ={}^1\mathrm{S}_0$ and $|e\rangle ={}^3\mathrm{P}_0$ \cite{2010GorshkovNatPhys}. Without breaking the $\mathrm{SU}({\cal N})$ symmetry, such a system allows for four distinct interaction strengths depending on the orbital states of the interacting fermions. Systems with this kind of two-orbital dynamics have been studied experimentally in 3D settings \cite{2014ScazzaNatPhys,2014ZhangScience}. 

The physics of such a two-orbital higher-spin system in the one-dimensional case has been theoretically explored in \cite{2013SzirmaiPRB} where the authors analyzed the case of atoms with $\mathrm{SU}(10)$ symmetry. At incommensurate filling, the phase diagram in the plane of different interaction strengths is quite intricate. Interestingly, the system presents the possibility of realizing a novel form of an FFLO state, where the finite momentum of the pairs does not come from the spin imbalance but rather from the difference of Fermi momenta of the two orbital states. 

Instead of using the two $|g\rangle = {}^1\mathrm{S}_0$ and $|e\rangle = {}^3\mathrm{P}_0$ states, an alternative way to realize a two-orbital system is to exploit the transverse single-particle modes of the trapping potential which is used to realize quasi-one-dimensional geometry. Specifically, if the atoms may occupy the first-excited degenerate states $p_x$, $p_y$ of the transverse potential, these states can play the role of the two orbitals \cite{2012KobayashiPRL,2014KobayashiPRA}. The phase diagram for a two-orbital fermionic system confined in a 1D lattice with incommensurate filling was explored theoretically in \cite{2016BoisPRB}, for both the ${}^1\mathrm{S}_0/{}^3\mathrm{P}_0$ and the $p_x/p_y$ two-orbital models. 

Under certain circumstances, two-orbital higher-spin systems mentioned above may support topologically nontrivial phases, including symmetry-protected ones. For example, as shown in \cite{2013NonneEPL}, the interplay between the orbital and nuclear spin degree of freedom for a one-dimensional optical lattice system at half-filling may lead to an interesting analogue of the Haldane phase \cite{2011CarrBook}. Creating symmetry-protected topological phases in $\mathrm{SU}({\cal N})$ 1D lattice systems was also explored in \cite{2015BoisPRB-Phase,2018UedaPRB}. This path of exploration is still ongoing and awaits experimental confirmation. 


\section{Conclusion}
\label{sec-concl}
It is a matter of fact that the one-dimensional many-body quantum systems are no longer only theoretical divagations. Due to the rapid experimental progress in controlling atoms and molecules in the ultra-cold regime, they become realistic systems having their own and very often exotic properties. It is highly possible that these unique features will find unconventional applications in the nearest future and will change many technological aspects. 

With this short review, we have summarized current progress in three directions which, in our opinion, are important not only from the point of technological exploitation but also have fundamental importance for developing our understanding of quantum many-body systems. Nonetheless, research in the field of one-dimensional fermionic systems includes many other topics which we have not discussed here. One example is systems with long-range interactions, such as dipolar gases \cite{2012BaranovChemRev}. The interesting effects of long-range interactions in one-dimensional dipolar Fermi gases have received considerable interest in the recent years \cite{2013DeSilvaPLA,2013XuPRL,2015MosadeqPRB,2015GrassPRA}. Another example is systems described by spin chain models, such as the Heisenberg spin model. Such models are closely associated with lattice systems, but have recently been considered as an effective description of strongly interacting one-dimensional atoms without a lattice potential \cite{2014DeuretzbacherPRA,2015YangPRA,2015LevinsenSciAdv,2015MurmannPRL}. Another interesting topic is the physics of one-dimensional Bose-Fermi mixtures, which have been recently studied theoretically \cite{2016HuNJP,2017DecampNJP,2017DeuretzbacherPRA,2017DehkharghaniJPB}.
 
Quantum simulators start to play a major role in various branches of physics. One can expect they will continue to be an increasingly versatile and useful tool. As an example of their potential, quantum simulators based on lattice models might help to understand the Holy Grail of modern condensed physics -- the high-temperature superconductivity. Although this phenomenon has been observed for many years, its underlying mechanisms are still not completely understood. Quantum simulators of high-temperature superconductor models may get us closer to unveiling this mystery \cite{2006KleinPRA}. Another example is the use of quantum simulators to study topological phases of matter. The current advances in generating artificial gauge fields enable major possibilities in this area.

\begin{acknowledgments}
This work is supported by the (Polish) National Science Center through Grant No. 2016/22/E/ST2/00555.
\end{acknowledgments}

\vspace{0.5cm}
\noindent{\bf Conflict of interest}

The authors declare no conflict of interest.

\vspace{0.5cm}
\noindent{\bf Keywords}

quantum simulators, one-dimensional systems, unconventional pairing, spin-orbit coupling, higher-spin systems

\bibliography{_Biblio}
    
\end{document}